\documentclass[
 reprint,
superscriptaddress,
 amsmath,amssymb,
 aps,
 pre,
floatfix,
]{revtex4-2}

\usepackage{graphicx}
\usepackage{dcolumn}
\usepackage{bm}
\usepackage{hyperref}

\usepackage[dvipsnames]{xcolor}

\usepackage{booktabs}

\usepackage{siunitx}
\usepackage{soul}
\soulregister\cite7
\soulregister\ref7
\soulregister\pageref7
\usepackage{ulem}

\newcommand{\Dphimax}{\Delta\phi_\text{max}}

\newcommand{\Rn}[1]{\mathbb{R}^{#1}}
\newcommand{\pbasin}{p_\text{basin}}
\newcommand{\ellipticK}{K}
\newcommand{\Rfp}[1]{R_{*#1}}
\newcommand{\rfp}[1]{r_{*#1}}
\newcommand{\params}[1]{\alpha_{#1},\delta_{#1},\epsilon_{#1},\omega_{#1}}

\newcommand{\tnl}{t_\text{NL}}
\newcommand{\Xs}{X_\text{S}}
\newcommand{\OmegaRD}{\Omega_\text{RD}}
\newcommand{\OmegaD}{\Omega_\text{D}}

\newenvironment{movetext2}{\color{orange}}{}
\newcommand{\eqnref}[1]{Eq.~\eqref{#1}}
\newcommand{\Equationref}[1]{Equation~\eqref{#1}}
\newcommand{\appref}[1]{Appendix~\ref{#1}}
\newcommand{\subfiglabel}[1]{{(#1)}}
\newcommand{\figref}[1]{Fig.~\ref{#1}}
\newcommand{\Figureref}[1]{Figure~\ref{#1}}
\newcommand{\subfigref}[2]{\figref{#1}\subfiglabel{#2}}
\newcommand{\secref}[1]{Sec.~\ref{#1}}

\newcommand{\comsolfull}{\texttt{COMSOL Multiphysics}}
\newcommand{\comsol}{\texttt{COMSOL}}
\newcommand{\subsectionref}[1]{Subsection~\ref{#1}}

\begin{document}

\preprint{APS/123-QED}

\title{Nonlinear stabilization of chiral modes in space-time modulated parametric oscillators}

\author{Scott Lambert}
\email{slamber5@uoregon.edu}
\affiliation{
 Materials Science Institute and Institute for Fundamental Science,
 University of Oregon, Eugene, OR 97403
}
\affiliation{
 Department of Physics, University of Oregon, Eugene, OR 97403
}
\author{Elise Jaremko}
\affiliation{
 Department of Physics, University of Oregon, Eugene, OR 97403
}
 \author{Jayson Paulose}
 \email{jpaulose@uoregon.edu}
\affiliation{
 Materials Science Institute and Institute for Fundamental Science,
 University of Oregon, Eugene, OR 97403
}
\affiliation{
 Department of Physics, University of Oregon, Eugene, OR 97403
}

\begin{abstract}

Phase control of parametric modulation in coupled oscillator networks enables
these networks to access dynamical states with desired spatiotemporal
symmetries. Floquet theory-based approaches successfully predict such
states in linear systems, but whether their symmetry properties persist under
nonlinearity remains largely unexplored. Here, we establish the existence of
nonlinear chiral steady states in a ring of three coupled parametric oscillators
with modulation phases chosen to selectively amplify a circulating mode in the
linearized system. We find that a cubic nonlinearity limits the exponential
growth of the amplified mode, producing a steady finite-amplitude motion that
retains the expected chirality. By exploiting the system's spatiotemporal
symmetry, we reduce the dynamics to a single averaged equation that
quantitatively predicts nonlinear trajectories, steady-state amplitudes, and
characteristic time scales of the full three-body system. The chiral steady
states possess finite basins of attraction and are accessible from wide ranges
of initial conditions and system parameters. Finite-element simulations of
elastic plate resonators quantitatively reproduce these features, establishing
the relevance of the reduced model to realistic continuum systems. Our results
demonstrate that desirable properties of linear time-modulated systems, such as
chirality and directional amplification, persist into strongly nonlinear
regimes, opening pathways to robust nonreciprocal signal routing and
amplification in parametrically driven platforms.
\end{abstract}

\maketitle

\section{Introduction}

Parametric oscillators, which are driven by time-modulation of system parameters
rather than direct forcing, represent a fundamental mechanism for mode-coupling
and amplification with broad applications such as low-noise amplification in
microwave~\cite{mumford2007some,smith1985low} and
superconducting~\cite{aumentado2020superconducting} circuits,
quantum~\cite{wu1987squeezed} and classical~\cite{rugar_mechanical_1991}
squeezed state generation, and coherent frequency conversion~\cite{louisell}.
Nonlinearities make new phenomena possible in these systems, such as the
stabilization of parametrically amplified modes~\cite{eichler2023classical},
hysteretic amplitude and phase jumps that can be used for sensitive
measurement~\cite{leuch2016parametric,rhoads2010impact,papariello_ultrasensitive_2016},
and state switching between stabilized phase states~\cite{leuch2016parametric}.
An analogy between the two stabilized phase states in a nonlinear parametric
oscillator (NPO) and the two states of a binary variable~\cite{Goto1959} suggests that coupled NPOs
could find the ground state of random Ising models, an NP-hard problem
\cite{wang2013coherent,Marandi2014,McMahon2016,heugel2022ising,calvanese2019theory,calvanese2021can}.
The spontaneous breaking of symmetry between the two phase states across coupled
NPOs can serve as a manifestation of a discrete time
crystal~\cite{heugel_classical_2019,Yao2020,yi2024theory}.

The theory of isolated NPOs is now well-established~\cite{eichler2023classical},
but the characterization of coupled systems of NPOs remains in development.
While many studies have investigated the dynamics of coupled NPOs with identical
in-phase parametric modulations occurring across all
oscillators~\cite{lifshitz2003response,Bromberg2006,lifshitz_nonlinear_2008,heugel_classical_2019,heugel2022ising,margiani2025three,Ameye2025}
which is relevant to Ising optimization and time crystallization, less is known
about coupled NPOs where the time-modulation has identical periodicity across
all oscillators (hence establishing a common time period for the system) but has
different phases on different oscillators. A phase offset in the parametric
modulation of a pair of NPOs influences the modulation frequencies at which
resonant steady states occur~\cite{Bello2019,calvanese2019theory}, but the
effect of phase variation on the resulting nonlinear trajectories was not
analyzed. The \textit{linear} analysis of rings of phase-shifted parametric
oscillators has predicted chiral~\cite{Downing2020,Melkani2024} and
one-way~\cite{Kruss2022,Melkani2025} wave transport and amplification with
potential applications in nonreciprocal response and
switching~\cite{Zangeneh-Nejad2019}. Whether these properties can be stabilized
in NPO circuits remains an open question, though recent work has identified self-oscillating states in phase-offset NPO rings in a distinct dynamical regime~\cite{Mahore2026a}.

\begin{figure}
  \centering
  \includegraphics{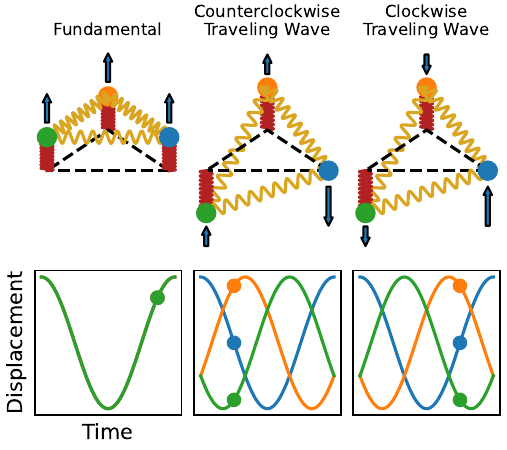}
  \caption{ The ``trimer" model and its normal modes. Each colored disc represents an
  identical mass. Each grounding spring (red) has a time-modulated linear component and a
  nonlinear component. The modulation phase of each grounding spring increases by
  $2\pi/3$ moving one position clockwise around the trimer (blue $\to$ green $\to$ orange). The coupling springs (gold) add a
  linear restoring force in the vertical direction. The positions and blue
  arrows represent the displacements and velocities of the masses from snapshots
  of the three normal modes of the linear system without modulation. The dots
  indicate the times and positions of the motion that is shown in the diagrams.
  }
  \label{fig:trimer}
\end{figure}

Here, we show that phase offsets in parametric modulation enable stable chiral
steady states in certain models of coupled
NPOs. Specifically, we investigate the dynamics of rings of three classical parametric
oscillators, representing the smallest network that can support a circulating mode with a well-defined chirality.
Without parametric modulation, the linear dynamics of the ring are completely described by its three normal modes, including two degenerate traveling-wave modes (\figref{fig:trimer}).
Prior work~\cite{Melkani2024} showed that applying a parametric resonance with phase shifts of $2\pi/3$ between adjacent oscillators breaks this degeneracy, singling out one of the traveling-wave modes for parametric resonance.
In the linear system, the resonant mode signaled an instability: an exponential growth of amplitude with time.
In this work, we establish that nonlinear restoring forces arrest this instability to yield a stable, finite-amplitude traveling wave that inherits the chirality of the resonant linear mode.

We demonstrate the stable chiral traveling-wave states in two complementary models.
We first investigate the effects of phase-offset parametric modulation in finite-element simulations of a ring of elastic plate resonators, as a demonstration of the proposed states in an experimentally realizable NPO system~\cite{Mestre2025}.
We find that the finite-element trajectories converge to persistent traveling-wave states, confirming their existence in a physically realistic model with thousands of degrees of freedom approximating a continuum system. 
We then recreate the same qualitative behaviors in a minimal discrete model with three coupled degrees of freedom, which incorporates the key ingredients of phase-offset parametric modulation and stiffening anharmonic response needed to generate the chiral steady states.
The minimal model, which we term the \emph{trimer}, is amenable to analysis through various approximation methods, which allows us to 
quantitatively account for and predict the characteristics of the steady states in the elastic plate model.
Since the plate resonator model incorporates all the essential ingredients of NPO networks---distinguishable oscillator degrees of freedom
that can be coupled, whose natural frequency (corresponding to the effective
stiffness) can be modulated in time, and which have 
a stiffening anharmonic response at large amplitudes---we expect the chiral stable states to be generally realizable in a variety of mechanical, optical, and electronic NPO systems~\cite{Mahboob2008,Mahboob2014a,Mathew2016,Seitner2017,Miller2019b,margiani2025three,Mestre2025}, with our minimal trimer model providing an analytically tractable description of the nonlinear chiral states across platforms.

This paper will proceed as follows. In Section \ref{sec:comsol}, we describe
finite-element simulations showing a stable, chiral oscillatory state arising from phase-offset parametric modulation
in a ring of three elastic plate resonators, 
thereby establishing the physical relevance of the phenomenon in advance of a detailed analysis.
Then, in \secref{sec:linear}, we present
the discrete trimer model equations of motion 
and its linear Floquet analysis.
In \secref{sec:nonlinear}, we derive
\textit{reduced averaged equations} for the nonlinear dynamics of the trimer, 
which enable a dynamical reduction to one degree of freedom that allow us to
analytically characterize the nonlinear chiral steady states. We show that the
reduced equations successfully predict the behavior of the continuum model with
parameters extracted independently from separate regimes of the finite-element
simulation dynamics. To the authors' knowledge, it has not previously been shown
that the dynamics of a coupled NPO system with nonuniform chiral modulation can
consistently converge to chiral motion described by a single degree of freedom,
with the effect of inter-oscillator coupling manifesting as a rescaling of the
single oscillator parameters.
We discuss the significance of this result and the connection to other known dynamical features of NPO networks in \secref{sec:discussion}.
We close with a discussion in \secref{sec:conclusions} of the implications of
our results for designing circulatory trajectories through phase modulation in
NPO networks and other time-modulated nonlinear dynamical systems.

\section{Simulation of chiral steady state in coupled plate
resonators}\label{sec:comsol}

\begin{figure*}[t]
  \centering
  \includegraphics{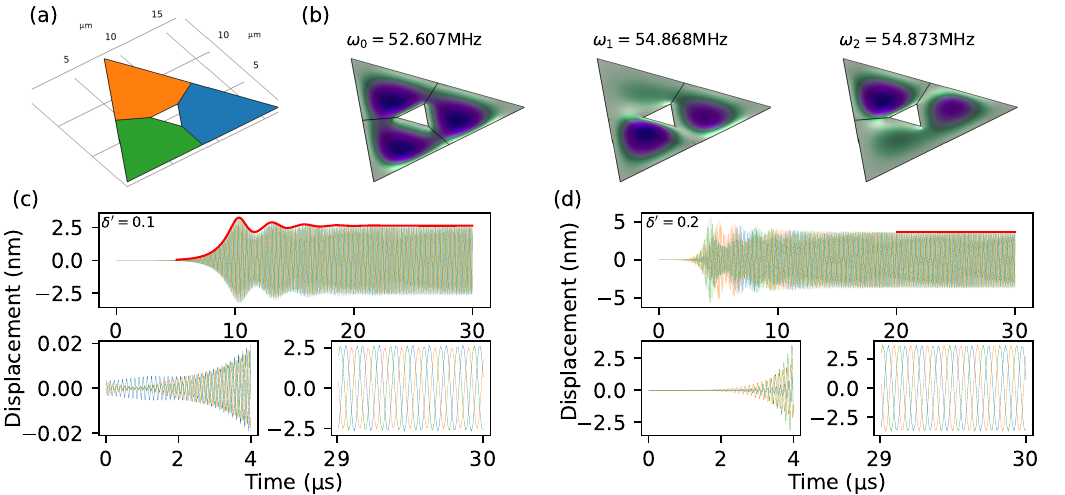}  
  \caption{Chiral steady states in a continuum model of mechanical parametric
    oscillators. (a) System of plate resonators arranged in a triangle. Black
    lines delineate three regions within which the in-plane stresses are
    modulated with the same phase offsets as in the discrete model (colors
    indicate indexing as in \figref{fig:trimer}, which dictates the phase offsets). (b) Eigenmodes of the
    static system with lowest three eigenfrequencies. The second and third
    eigenfrequencies differ by a fraction of less than $10^{-4}$. The next
    eigenfrequency is well-separated at $\omega_3 = 76.67 \, \mathrm{MHz}$. (c)
    Numerical time-evolution of vertical displacements at the centroids of the
    three resonators under space-time modulation of in-plane stresses with
    $\delta' = 0.1$. Lower panels show zoomed regions at the initial and final
    stages of the simulation. Red curve shows numerical integration of the
    reduced averaged equations derived in \secref{sec:nonlinear}, with parameters extracted as described in
    \appref{app:comsolfit}. (d) Thin curves: same as (c) but at a higher
    modulation strength $\delta' = 0.2$. Thick red line shows predicted
    steady-state amplitude from the discrete trimer model with twice the
    modulation strength as (c), see \appref{app:comsolfit}. }
  \label{fig:comsol}
\end{figure*}

We begin by demonstrating nonlinear-stabilized chiral parametric oscillations in a physically realistic model of coupled elastic plate resonators~\cite{Mathew2016,Prasad2017,Naserbakht2019,Su2021}.
The out-of-plane deflections of thin elastic plates experience restoring forces that are inherently anharmonic at large deflections due to geometry~\cite{ventsel}, without requiring material nonlinearities.
The geometric nonlinearity also couples out-of-plane deflections to in-plane prestresses, providing a mechanism for parametric modulation of vibrations using electrostatic~\cite{Cha_2018,Mei_2018,Mestre2025} or thermal~\cite{Blaikie_2019} means.

We recreate the trimer configuration---the minimal assembly that supports chiral
modes---by arranging three triangular resonators in a ring with overlapping
corners [\subfigref{fig:comsol}{a}]. The assembly is modeled as a thin plate
with isotropic linear elasticity and clamped boundary conditions along the
external and internal edges (see \appref{app:comsol} for details of the model
and its implementation using finite elements). The lowest three modes in a
linear eigenmode analysis of the assembly can be interpreted as coupled modes
arising from the lowest-frequency continuum eigenmode on each resonator
[\subfigref{fig:comsol}{b}]. The lowest-frequency coupled eigenmode involves all
resonators vibrating in-phase, consistent with the fundamental normal mode of
the trimer (\figref{fig:trimer}). The next two modes are nearly degenerate and
involve out-of-phase superpositions of the individual resonator modes. These can
be interpreted as arbitrary orthogonal modes spanning the traveling-wave
degenerate normal mode subspace of the trimer, with the degeneracy between the
clockwise and counterclockwise modes lifted by the randomly generated
finite-element mesh, which breaks the exact threefold rotational symmetry of the
continuum model. 

The natural frequency of the individual resonator modes can be varied by modulating the in-plane tension, akin to tuning a drum~\cite{ventsel}; this mechanism is routinely used for parametric excitation of plate resonator modes~\cite{Bachtold2022}.
The  strongest parametric resonances arise when the tension is modulated at close to twice the eigenfrequency of a normal mode~\cite{landau1960mechanics}.
However, a uniform modulation of the in-plane tension at $2\omega_1 \approx 2\omega_2$ does not discriminate between the two counterpropagating modes; to do so, the modulation must itself break the chiral symmetry of the assembly.
According to the linear analysis of Ref.~\onlinecite{Melkani2024}, a chiral parametric resonance is obtained by introducing a modulation phase offset of $2\pi/3$ between adjacent degrees of freedom.
To achieve this, we subject each resonator, indexed by $j \in \{1,2,3\}$, to a uniform
in-plane bulk stress with time dependence
\begin{equation}
  \label{eq:membranetension}
  \sigma_{xx}(t) = \sigma_{yy}(t) = \sigma_0\left[1+\delta' \cos \left(\Gamma t + \frac{2\pi}{3}j  \right) \right],
\end{equation}
where $\sigma_0$ is a uniform isotropic prestress modulated at frequency $\Gamma$ with strength $\delta'$.
The chosen phase offset is expected to parametrically excite the counterclockwise traveling wave for modulation frequencies at or near twice the traveling-wave normal mode frequency~\cite{Melkani2024}.

To test the linear expectation in the continuum model and investigate the subsequent nonlinear response at large deformations, we performed finite-element simulations on resonance ($\Gamma = 2 \omega_1$). 
The effect of damping in the system was recreated by adding a
dissipative component to the elastic modulus (\appref{app:comsol}). Since the numerical eigenfrequency
analysis of the plate system provides an arbitrary linear superposition of the
degenerate clockwise and counter-clockwise propagating modes,
 we are not able to initialize the system
purely in the resonant mode. Instead, we initialized the system in the second
linear eigenmode, ensuring some overlap of the initial condition with the
desired counterpropagating mode. We tracked the vertical displacements of the
centroids of the three colored regions as proxies for the states of the
fundamental mode oscillations within each plate resonator.

Results of a dynamical simulation with space-time modulation strength $\delta' = 0.1$ are shown in \subfigref{fig:comsol}{c}. At this modulation, the system
shows an initial transient around \qty{2}{\micro\second} long, after which the
three resonators oscillate at roughly $2\pi/3$ out of phase, forming a
counterclockwise traveling wave. 
Subsequently, the amplitude of the wave increases faster-than-linearly with time over dozens of oscillation cycles, consistent with the expected parametric resonance.
Under linear dynamics, the resonant mode would grow exponentially over time without bound.
However, the geometric nonlinearity in the elastic restoring forces experienced by the plate becomes significant as the vibration amplitude increases, eventually arresting the growth at $t \approx \qty{10}{\micro\second}$.
Following a period  of slow variations in the wave amplitude, reminiscent of amplitude ringdown in damped oscillators, the trajectory settles into a constant-amplitude steady state that maintains the counterclockwise sense in successive oscillation peaks of the traveling wave.

The numerical trajectory presented in \subfigref{fig:comsol}{c} demonstrates that a chiral nonlinear steady state can be achieved via phase-offset parametric modulation by using insights from linear instability analysis. 
The geometric nonlinearity provides stability without disrupting the spatiotemporal structure of the initial resonant mode.
However, this demonstration does not address under what conditions the desired steady state can be accessed.
In a linear system with parametric resonances, the unstable mode with the fastest growth rate always dominates the signal at long enough times, since its amplitude grows exponentially faster than any other mode with time. 
In the nonlinear system, by contrast, the exponential growth occurs within a finite time, so the dominance of the counterclockwise traveling-wave mode is no longer guaranteed if the system is not initialized precisely in that mode as in our simulations.

To investigate whether the chiral steady state requires the unstable mode to
dominate during the exponential growth phase, we performed a finite-element
simulation with modulation strength  $\delta' = 0.2$
[\subfigref{fig:comsol}{d}]. At this stronger pumping, the initial evolution
from the arbitrary member of the degenerate traveling-wave mode subspace quickly
deviates from the counterclockwise traveling-wave mode, favoring one resonator
over the other two. However, the resonator amplitudes begin to even out during
the ringdown phase (\qty{5}{\micro\second}--\qty{15}{\micro\second}) which ends
with the system again reaching a chiral steady state with the phase offsets
corresponding to the counterclockwise traveling mode. This simulation shows that
the chiral steady state can be reached at late times even if the trajectory
deviates from the unstable traveling wave at early times, 
suggesting a degree of robustness to the chiral steady state in the form of a finite basin of attraction.

While the finite-element simulations establish the existence and experimental relevance of nonlinear chiral steady states generated by parametric resonances, the nonlinear plate equations describing the continuum system are challenging to analyze directly.
To place the qualitative observations of this section on a rigorous footing, we instead perform an in-depth analysis of the discrete coupled oscillator model illustrated in \figref{fig:trimer}, augmented with the key ingredients of nonlinear restoring forces and phase-offset parametric modulation. 
The discrete model, which we refer to as the trimer, provides a tractable description of the complex continuum dynamics when restricted to the relevant subspace of coupled primary displacements of individual resonators.
Despite the drastic reduction in degrees of freedom from thousands of mesh elements to three, the trimer quantitatively reproduces many features of the plate resonator trajectories 
(red curves in \figref{fig:comsol}), 
providing an efficient description of the physical mechanisms responsible for the chiral steady states.

\section{Trimer model: Equations of motion and linear behavior} \label{sec:linear}

The trimer model describes the dynamical evolution of three coupled oscillators arranged as in \figref{fig:trimer} through the non-dimensionalized equations of motion
\begin{align}\label{eq:trimereom}
    \ddot{x}_j + \left[1+\delta\cos(\gamma t + \beta_j)\right]x_j + \epsilon \dot{x}_j 
    + x_j^3 \\+ k(2x_j - x_{j+1} - x_{j-1}) = 0. \nonumber
\end{align}
Here,  $j \in \{1,2,3\}$ indexes the oscillators, and $x_j$ is the displacement
of oscillator $j$ from equilibrium (with wraparound: $x_4 = x_1$ and $x_0 =
x_3$). 
In addition to a harmonic restoring force with a phase-shifted parametric modulation on each site, the model incorporates linear damping, a cubic nonlinearity in the restoring force, and a static coupling among adjacent oscillators.
The dimensionful equations and the non-dimensionalization procedure are
presented in \appref{app:nondim}. We have chosen the time unit so that the
frequency of each oscillator (in isolation and without parametric
modulation) is unity, and the length unit so that the cubic nonlinearity has
unit coefficient. The parameters $\delta$ and $\gamma$ respectively set the
fractional strength and frequency of the parametric modulation. The phase offset $\beta_j$
is set to
\begin{equation} \label{eq:betas}
  \beta_j = \frac{2\pi}{3}(j-1)
\end{equation}
to generate a traveling-wave modulation. 
The parameter $\epsilon$ sets the strength of the linear damping, and $k$ is the coupling
between adjacent oscillators. 
The linear eigenmodes of the system with $\delta=\epsilon=0$ and ignoring the
cubic term are shown in \figref{fig:trimer}. The corresponding eigenfrequencies
are $\omega=1$ for the fundamental mode and $\omega = \sqrt{1+3k}$ for the
degenerate traveling modes.

\begin{figure}
    \centering
    \includegraphics{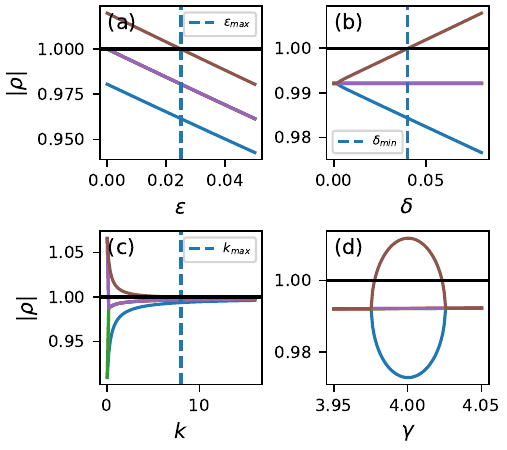}
    \caption{Floquet spectra of the trimer at various parameter values. The
    colored lines are the absolute values of the Floquet eigenvalues of the
    system at the value of the parameter shown on the horizontal axis. When an
    eigenvalue exceeds $|\rho|=1$ (black lines), amplification occurs. The
    vertical lines indicate values where the fixed point formula
    \eqnref{eq:trimerfp} predicts amplification will no longer occur. For the
    plots where $\epsilon$, $\delta$, and $\gamma$ are varied, the other parameters
    are held fixed; the fixed values are $k = 1$, $\epsilon = 0.01$, $\delta =
    0.1$, and $\gamma = 2\omega = 4$. For the plot with varied $k$, the modulation frequency is also varied to
    maintain the resonance condition $\gamma = 2\omega = 2\sqrt{1+3k}$. }
    \label{fig:baselinespectra}
\end{figure}

In the absence of nonlinearity, the time-modulated system can be analyzed using
Floquet theory~\cite{YakubovichStarzhinskii,Melkani2024}: the phase-space
trajectory of the system can be written as a linear combination of six
\textit{Floquet modes} $\vec\psi_i(t)$. Each Floquet mode is a 6-vector of three
displacements and three velocities; the six modes span the phase space of the
trimer. \textit{Floquet's theorem} says that over one modulation period $T$,
each Floquet mode changes by a constant multiplier: $\vec\psi_i(t+T) = \rho_i
\vec\psi_i$. The multiplier $\rho_i$ is the \textit{Floquet multiplier} of the
$i$th Floquet mode and is in general complex. Floquet modes with $|\rho_i| \neq
1$ signify exponentially growing or decaying modes. At finite damping, some
Floquet modes decay over time ($|\rho_i| < 1$). However, at particular
modulation parameters, some Floquet multipliers may have magnitude greater than
one, signifying \textit{parametric resonances} that feed energy into the system
over many cycles.

The specific choice of modulation phases, \eqnref{eq:betas}, endows the system
with a space-time symmetry~\cite{golubitsky2002,Xu2018}: the equations of motion are unchanged upon
simultaneously advancing time by $T/3$ and increasing the oscillator coordinate
index by one~\cite{Melkani2024}. This
symmetry constrains the parametric resonances of the system: when the modulation
frequency is set to twice the frequency of the traveling-wave normal modes,
$\gamma = 2\omega$, only one mode is parametrically amplified, which smoothly
connects to the counterclockwise normal mode as $\delta \to
0$~\cite{Melkani2024}. At zero damping, this mode is accompanied by an
exponentially decaying Floquet mode which also rotates counterclockwise, but is
offset by $\pi/2$ in its phase relative to the amplified mode. 
The other four Floquet modes, which derive from the fundamental mode and the
clockwise traveling wave mode, are not resonances but are steady oscillations with
$|\rho|=1$ at zero damping.

\figref{fig:baselinespectra} shows the dependence of the trimer Floquet
multipliers on the system parameters over representative ranges; in particular, it shows values of the
parameters where amplification does and does not occur. Brown and blue curves
show the parametrically amplified and attenuated modes respectively, while other
colors show the other modes. In \subfigref{fig:baselinespectra}{a}, the modes
shown in purple coincide at $|\rho|=1$ when $\epsilon=0$, as expected of modes
that are not parametrically resonant. As damping increases, it decreases the
magnitudes of the Floquet multipliers, but the single amplified mode continues
to be amplified until a threshold damping value (marked $\epsilon_\text{max}$)
beyond which $|\rho|$ drops below one and all modes are attenuated.
\subfigref{fig:baselinespectra}{b} shows the amplification rate of the amplified
mode growing with the strength of the modulation; the leading behavior is linear at the small $\delta$ values shown here but becomes more complex at larger modulation strengths~\cite{kovacic2018mathieu}. 
\subfigref{fig:baselinespectra}{c} shows that increasing the coupling strength
while keeping other parameters constant (but increasing the modulation frequency 
to maintain the resonance condition) leads to a reduction in the degree of
amplification; this can be interpreted as a result of the modulation
only acting on the grounding springs. 
Finally, \subfigref{fig:baselinespectra}{d} shows the finite range of modulation
frequencies that result in amplification. The parametric resonance is predicted
to occur exactly at $\gamma = 2\omega$ in the $\delta \to 0$ limit, but at
finite modulation strengths, the resonance broadens to a range of modulation frequencies around the primary value, albeit at lower strengths.
 While all these
resonance conditions can be derived using Floquet theory, we will recover them
in our averaging analysis of the full nonlinear equations of motion in
subsequent sections. 

The linear theory of the trimer, and in particular its
space-time symmetry, singles out the counterclockwise mode for parametric
amplification, which occurs for finite ranges of the parameters and does not
require fine tuning. Regardless of initial condition, we expect the linear
dynamics to eventually be dominated by the resonant mode, which grows
exponentially with time. Therefore, the system will eventually reach a regime
where the nonlinear term becomes significant, regardless of the size of the
initial displacements. In the next section, we address the change in the
long-time behavior of the amplified mode due to the nonlinearity. We will find
that the nonlinearity limits the exponential growth of the amplified mode by
shifting the phase of the system's oscillations relative to the parametric
drive, causing the amplitude of all three oscillators to converge to a fixed
value.

\section{Nonlinear dynamics of amplified chiral mode}\label{sec:nonlinear}

\begin{figure}
    \centering
    \includegraphics{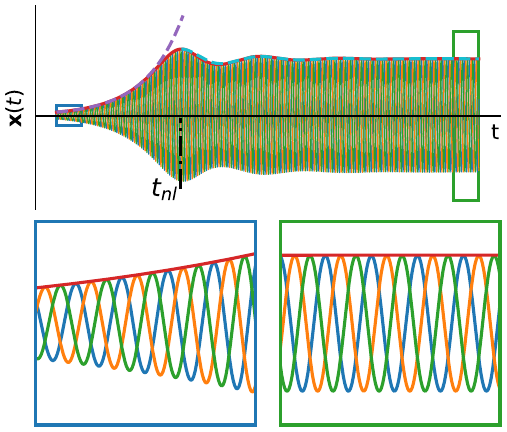}
    \caption{Numerical solution of the full equations of motion. The parameters
         are $\epsilon = 0.06$, $\delta = 0.6$, $k = 1$, $\omega = \sqrt{1+3k} =
         2$, $\gamma = 2\omega = 4$. 
         The system is initialized in its unique amplified Floquet eigenstate,
         normalized to have magnitude 0.1. Solid red curve shows the amplitude
         obtained by solving the reduced averaged equations
         \eqref{eq:realreducedeomresonance}. Blue boxes indicate the extents of the
         insets below. \textbf{Top:} The full trace from $t = 0$ to $t = 250$.
         Three analytical predictions described in \secref{sub:avgdyntimescales} are indicated:
         Dashed purple curve shows exponential growth with rate $g$ predicted in \eqnref{eq:growthrate}; vertical dash-dotted line shows the time $\tnl$ marking the transition to nonlinear behavior; cyan blue curve shows ringdown modeled as a damped harmonic oscillator. \textbf{Blue inset:}
         The initial exponential amplification. \textbf{Green inset:} The steady
         state showing chiral motion (each coordinate oscillating out-of-phase
         by $2\pi/3$). Both insets span the same time interval. 
         }
    \label{fig:threepanel}
\end{figure}

We now investigate the dynamics of the trimer by numerically integrating the
full equations of motion [\eqnref{eq:trimereom}], with parameters chosen to lie
in a regime where an amplified chiral mode is predicted by the linear theory.
\Figureref{fig:threepanel} shows a numerically integrated trajectory for such a
system, initialized in the amplified Floquet mode. Consistent with the linear
theory, the mode amplitude grows exponentially at early times, with a rate
predicted by the magnitude of the corresponding Floquet multiplier. However, the
trajectory eventually deviates from the linear behavior, displaying oscillations
in the amplitude that decay to attain a steady amplitude at late times. The
insets to \figref{fig:threepanel} show that the relative phases of the three
oscillators in the steady state match the initial phases, confirming that the
chirality of the parametrically amplified Floquet mode is preserved in the
nonlinear-stabilized steady state.

The qualitative features of the trajectory
described above---early exponential growth of the counterclockwise traveling wave, followed by decaying amplitude oscillations to a chiral steady state---are observed for a wide range of trimer parameters.
Notably, the same features are present in the plate resonator trajectory of \subfigref{fig:comsol}{c} after the initial transient, suggesting that they are not model-specific and may serve to broadly characterize the evolution of the unstable chiral mode from the linear into the nonlinear regime in a variety of coupled NPO architectures.
In the following subsections, we show that an averaging method reduces the coupled trimer dynamics to a single equation for the amplitude of the chiral mode, yielding quantitative predictions for the
steady-state amplitude and the time scales of the exponential growth, the
appearance of the amplitude oscillations, and their eventual decay as functions of the parameters in \eqnref{eq:trimereom}.
We further show that these parameters can be extracted directly from continuum simulations (\appref{app:comsolfit}), establishing the utility of the trimer model as a quantitative design tool for chiral steady states in NPO platforms.

Throughout this section, numerical integrations follow procedures described in
\appref{app:methods}.

\subsection{Reduced averaged equation: reducing the coupled system to a single degree of freedom}

In order to understand the long-term dynamics of the trimer model, we use
averaging to reduce the equations of motion \eqref{eq:trimereom} to an
approximate autonomous system
\cite{strogatz_nonlinear_1994,sanders_averaging_2007,guckenheimer1983nonlinear}.
This produces results equivalent at first order in the system parameters to the
two-variable expansion method of
\cite{kovacic2018mathieu,strogatz_nonlinear_1994}, and has been applied
previously to analyze slow dynamics of parametric oscillators
\cite{eichler2023classical,lifshitz_nonlinear_2008,Bello2019}; it is accurate as
long as the ``slow'' variables evolve much more slowly than the system's
oscillation frequency. In this case, the relevant slow variables are the
amplitude and phase of the trimer oscillations, and the relevant ``fast''
frequency scale is the traveling-wave frequency, $\omega =
\sqrt{1+3k}$.\footnote{We do not prove this here, but if the fixed point
\eqnref{eq:trimerfp} is asymptotically stable, the error between the trajectory
of the averaged trimer equations of motion and that of the exact equations will
be bounded for all time. See \cite{sanders_averaging_2007}, Section~5.5, for
details.}

We encapsulate the ``slow'' amplitude and phase variables in a single complex
amplitude $A_j$ for each oscillator:
\begin{align}
    x_j &= A_j e^{i\omega t} + A_j^* e^{-i\omega t} \label{eq:complexcoordspos} \\
    \dot{x}_j &= i\omega A_j e^{i\omega t} -i\omega A_j^* e^{-i\omega t} \label{eq:complexcoordsvel} 
\end{align}
Making this coordinate change in the equations of motion and time-averaging over
one period while treating the amplitudes $A_j$ as constant results in an
autonomous equation of motion for each $A_j$. Substituting the ansatz 
\begin{align} \label{eq:ansatz}
    A_j = Ae^{-2\pi i (j-1)/3}
\end{align}
into these equations makes all three equations identical. The resulting
\textit{reduced averaged equation} for the trimer, derived in \appref{app:raederivation}, is 
\begin{align}\label{eq:reducedeom}
    \dot{A} = \frac{\delta}{4\omega}iA^*e^{i\Delta\gamma t}
    -\frac{1}{2}\epsilon A + \frac{3}{2\omega}i|A|^2A
\end{align}
where $\Delta\gamma = \gamma - 2\omega$ is the detuning from the resonance. 
 \Equationref{eq:reducedeom} describes the amplitude
and phase of the chiral motion through a single complex amplitude $A(t)$.
Although it describes the motion of all three oscillators at long times via the
ansatz \eqref{eq:ansatz}, it has the form of the averaged equations of motion of
a single nonlinear Mathieu oscillator with damping $\epsilon$, parametric drive
$\delta/\omega$, and nonlinearity parameter $1/\omega$ (see 
\appref{app:nondim} for an explanation of the nonlinearity parameter). 

Note that although the equations of motion have been reduced to a single degree
of freedom, the effect of the coupling $k$ is still present through the
traveling-wave frequency $\omega = \sqrt{1+3k}$, which effectively rescales the
parametric drive term and the nonlinearity. While equations of this form are
known in the literature \cite{lifshitz_nonlinear_2008,eichler2023classical}, the
authors are not aware of work showing that the motion of a coupled
system with chiral modulation converges to the dynamics of this single equation,
or how the coupled nature of the full system remains manifested in the reduced
equation via rescaling of parameters.

The solid line in \figref{fig:threepanel} shows a trace of $|2A(t)|$, the real
amplitude of the oscillations, obtained by numerically integrating the reduced
equation using the same parameters as the full dynamics. We observe that the
reduced equation solution closely follows the envelope of the full dynamics, and
captures all the salient features of the amplitude variation over time. 

To understand the behavior of the traces in \figref{fig:threepanel}, it is
useful to express the reduced averaged equations \eqnref{eq:reducedeom} in real
polar coordinates. Letting $r$ be the real amplitude of the traces and $c$ be
the phase of oscillator 0, the complex amplitude is expressed as
\begin{align}\label{eq:rctocomplexamp}
    A = \frac{r}{2}e^{ic}
\end{align}
and the reduced averaged equation \eqnref{eq:reducedeom} is recast as a pair of evolution equations
\begin{align}\label{eq:realreducedeom}
    \dot{r} &= \frac{\delta}{4\omega} r \sin(2c-\Delta\gamma t)-\frac{1}{2}\epsilon r \\
    \dot{c} &= \frac{\delta}{4\omega} \cos(2c-\Delta\gamma t) + \frac{3}{8\omega} r^2. \nonumber
\end{align}
The term $3r^2/(8\omega)$ originates from the nonlinearity in the original
equations of motion. When it is small, the amplitude experiences exponential
growth as predicted by Floquet theory. But as the amplitude increases, this
nonlinear term causes an increase in the phase growth rate $\dot{c}$, and this
increase grows with the amplitude. As such, the growth of the amplitude under
parametric amplification eventually shifts the phase of the oscillations such
that the parametric drive begins to remove energy from the system, rather than
adding energy. This in turn causes the amplitude to decrease, shifting the phase
and allowing the parametric drive to once again add energy, which causes another
phase shift and decrease in energy, and so on until the amplitude converges.
This is a physical explanation for the ringdown and convergence of the amplitude
shown in \figref{fig:threepanel}; we discuss these phases of the motion in the
following subsections.

\subsection{Time scales of the motion} \label{sub:avgdyntimescales}

We now use the reduced averaged equation \eqnref{eq:reducedeom} to predict the time scales
of the growth, transition, and ringdown phases identified previously in the full
dynamics of \figref{fig:threepanel}. While \eqnref{eq:reducedeom} and the results that follow explicitly include dependences on the frequency and strength of the parametric modulation as well as the damping, the dependence on the coupling constant $k$ is implicitly captured in the traveling-wave frequency
$\omega = \sqrt{1+3k}$, which encodes the effect of coupling in a single degree
of freedom through rescaled effective parameters.

\subsubsection{Initial amplitude growth rate}
Upon initializing the system in the resonant Floquet mode at a small amplitude, the linear dynamics predicts that the amplitude grows exponentially, $\sim e^{gt}$,
where the growth rate $g$ is determined by the Floquet multiplier. 
We can extract the value of $g$ from the small-amplitude limit of the reduced averaged equations, with the result
\begin{align}\label{eq:growthrate}
    g = \frac{1}{2}\left(\frac{\delta}{2\omega} - \epsilon \right)
\end{align}
(see \appref{app:nlderivations} for the derivation). This
growth rate matches the rate obtained in perturbative analyses of the linear
Mathieu equation \cite{kovacic2018mathieu,landau1960mechanics}.
Since the oscillation frequency $\omega$ grows with $k$, this expression shows that the growth rate decreases as the coupling grows, and that amplification can vanish at large enough coupling, consistent with the behavior of the numerically-obtained Floquet multipliers in \subfigref{fig:baselinespectra}{c}.
The dashed purple curve in \figref{fig:threepanel} shows that \eqnref{eq:growthrate} correctly reproduces the initial amplitude growth of the full nonlinear trajectory.

\subsubsection{Transition to nonlinear behavior}

We estimate the time scale at which the exact solution begins to deviate from
the exponential growth of the linear system by equating the amplitude of the
exponentially growing linear solution to a length scale $\Xs$ derived from the
nonlinearity. Upon initializing the system in the resonant Floquet mode with amplitude $a_0$ set to be small enough for the response to be effectively linear, the amplitude evolves as
\begin{align}
    a(t) = a_0 e^{gt},
\end{align}
with the growth rate $g$ from \eqnref{eq:growthrate}, and reaches the value $\Xs$ at the time 
\begin{align}\label{eq:nonlinearttgeneral}
    \tnl \equiv \frac{1}{g}\ln\frac{\Xs}{a_0}.
\end{align}
For $\Xs$, we use the dimensionful nonlinearity parameter $\alpha$ derived from
the dimensionful equations of motion. Before
nondimensionalization, the nonlinearity in the equation of motion for oscillator
$j$ has the form $\alpha x_j^3$, where $\alpha$ is a constant with dimensions of
$(\text{length})^{-2}$. We therefore set 
\begin{align} \label{eq:nonlinearlength}
    \Xs = \frac{1}{\sqrt{\alpha}}.
\end{align}
See Appendix \ref{app:nondim} for a detailed description of the
nondimensionalization procedure and the parameter $\alpha$.

The time scale $\tnl$, with growth rate and nonlinear length calculated using \eqnref{eq:growthrate} and \eqnref{eq:nonlinearlength} respectively, is indicated in \figref{fig:threepanel} as a vertical line, and matches the interval at which the nonlinear trajectory deviates strongly from the initial exponential growth. 
\eqnref{eq:nonlinearttgeneral} shows that the transition time depends on the
coupling through the growth constant $g$. In the case of the trimer,
\eqnref{eq:growthrate} and \eqnref{eq:nonlinearttgeneral} show that the
transition time grows as the coupling increases.

\subsubsection{Amplitude ringdown}

Once the amplitude stops growing exponentially, it undergoes decaying
oscillations around a fixed point. The dynamics of these oscillations can be
mapped to a damped harmonic oscillator by linearizing around the fixed point
(see \appref{app:nlderivations}).
For the trimer, the resulting damped oscillations have a natural frequency of 
\begin{align}\label{eq:linearizationfreq}
    \OmegaRD = \sqrt{\left(\frac{\delta}{2\omega}\right)^2-\epsilon^2}    
\end{align}
and damping parameter $\epsilon$.
This equation shows that for the trimer model, increasing the coupling
decreases the ringdown frequency of the damped oscillations. The dashed cyan
curve in \figref{fig:threepanel} shows a damped oscillation obeying
\eqnref{eq:bphase}, shifted to begin at the peak of the amplitude oscillations
and having an amplitude equal to the difference between the peak amplitude and
the fixed point amplitude.

A direct consequence of the mapping to a damped harmonic oscillator is that the ringdown lasts over a time scale of order $1/\epsilon$, which becomes longer as the damping is reduced and diverges in the undamped limit $\epsilon \to 0$.
This divergence is a signature of persistent amplitude oscillations in the undamped limit.
The reduced averaged equations with $\epsilon = 0$ admit a constant of motion
that allows the period of these oscillations to be expressed in terms of
elliptic integrals; these calculations and their verification in numerical
trimer trajectories are reported in \appref{app:undamped}. 

For systems with damping, the nonlinear trajectory settles into a traveling-wave steady state with a constant amplitude following the ringdown phase, which forms the focus of our work. 
We characterize this steady state using the reduced averaged equations in the following subsection.

\subsection{Fixed point formula for the chiral steady state}

\begin{figure}
    \centering
    \includegraphics{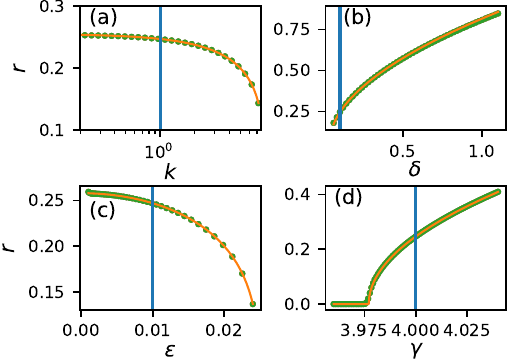}
    \caption{Comparison of steady-state amplitudes of numerical traces with
      the analytical  prediction from averaging theory.
      Numerically-computed amplitudes from full trimer trajectories are shown as dots. 
    In each panel, one parameter is varied while the other parameters are kept
    fixed at the baseline values indicated by vertical blue lines: $\epsilon =
    0.01$, $\delta = 0.1$, $k = 1$, $\gamma = 2\omega = 2\sqrt{1+3k} = 4$.
    Orange curves shows the prediction from averaging theory for the amplitude,
    \eqnref{eq:fixedamp}.
    Away from resonance, \eqnref{eq:minshift} predicts a finite-amplitude fixed point for $\gamma > 3.97708$.
  }
    \label{fig:4panel}
\end{figure}

\begin{figure}
    \centering
    \includegraphics{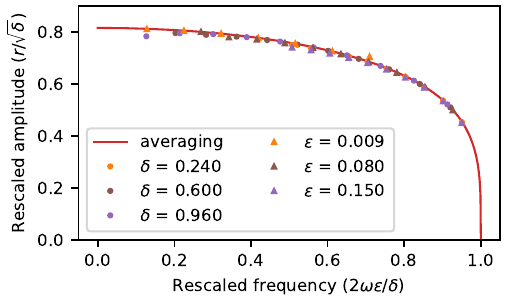}
    \caption{ Rescaled trimer steady-state amplitude from numerical integration
      of the full dynamics (symbols) as a function of rescaled frequency. Under
      these rescalings, the fixed-point amplitude predicted by averaging theory,
      \eqnref{eq:trimerfp}, becomes a universal function (solid curve).
      Different markers correspond to different combinations of the parameters
      $\delta$ and $\epsilon$. For the markers with varied $\delta$, $\epsilon$
      is fixed at 0.06, and for the markers with varied $\epsilon$, $\delta =
      0.6$. For each $(\delta,\epsilon)$ combination, the rescaled frequency is
      varied by changing $k$.
      }
    \label{fig:reducedplot}
\end{figure}

To quantify the constant amplitude attained by the resonant mode at long times,
we seek fixed points to the reduced averaged equations for the amplitude and
phase evolution. We first consider modulation exactly at resonance, with
modulation frequency set to twice the traveling-wave mode frequency $\omega$.
Upon setting the detuning $\Delta\gamma = 0$, the reduced equations
\eqnref{eq:realreducedeom} admit the fixed point $(r_*,c_*)$ given by
\begin{align}\label{eq:trimerfp}
    \sin{2c_*} = \frac{2\omega\epsilon}{\delta},\text{ } 
    r_* = \sqrt{\frac{2}{3}}\left[\delta^2-(2\omega\epsilon)^2\right]^{1/4}
\end{align}
This is the only stable fixed point of the reduced averaged equations with a
nonzero amplitude. 
Numerical investigation of the phase portraits generated by
\eqnref{eq:realreducedeom} suggested that all trajectories in the
reduced system eventually reach the unique fixed point when there is positive
nonzero damping. We hypothesize that all initial conditions in the 2D
phase plane of the reduced averaged equations eventually flow to the fixed point
so long as $r(0) > 0$. 
This suggests that in the \textit{full trimer dynamics}, 
systems initialized in the amplified traveling wave will
flow to an attractor corresponding to the fixed point
of the reduced averaged equations. 

To test this hypothesis, we numerically
integrated the full dynamical equations of the trimer after initializing the
system in the amplified Floquet mode with an overall magnitude of 0.1 while
varying system parameters (see Appendix \ref{app:methods} for procedural
details). On resonance ($\Delta \gamma = 0$), the full dynamics converged to the
predicted fixed point over wide ranges of system parameters, as shown by
comparing the amplitude extracted from numerical trajectories (symbols) to the
prediction of \eqnref{eq:trimerfp} (curves) in \figref{fig:4panel}(a--c).
The form of \eqnref{eq:trimerfp} shows that the dependence on the various
parameters at resonance can be captured by a single rescaled frequency variable
$\omega_\text{rs} = 2\omega\epsilon/\delta$, provided the amplitude is rescaled
as $r_\text{rs} = r_*/\sqrt{\delta}$. We confirm this universal dependence in
\figref{fig:reducedplot}, where we show that numerically-computed steady-state
amplitudes for a variety of parameters collapse onto a single curve $r_\text{rs}
= \sqrt{2/3}(1-\omega_\text{rs}^2)^{1/4}$. 
We also verified that the
trimer oscillation phases match the expectation of consistent $2\pi/3$ offsets
(\appref{app:chirality}).

Away from resonance---i.e., when $\Delta\gamma \neq 0$---the reduced equation no
longer admits a fixed point. However, it does admit a state in which the
amplitude $r$ is fixed and the phase offset $c$ changes linearly with
time~\footnote{Under the coordinate change $\tilde{r} = r$, $\tilde{c} = c -
\Delta\gamma t/2$, the constant-amplitude state is a true fixed point of a
variant of the reduced averaged equations \eqnref{eq:realreducedeom}:
$\dot{\tilde{r}} = \frac{\delta}{4\omega}r\sin(2\tilde{c})-\frac{1}{2}\epsilon
r$, $\dot{\tilde{c}} =
\frac{\delta}{4\omega}\cos(2\tilde{c})+\frac{3}{8\omega}r^2-\frac{\Delta\gamma}{2}$}.
Setting $c = c_0+c_1t$, substituting into \eqnref{eq:reducedeom}, and setting
$\dot{A} = 0$, we obtain
\begin{align}\label{eq:fixedampphase}
    c_1 = \frac{\Delta\gamma}{2}, \text{ }
    \sin(2c_0) = \frac{2\omega\epsilon}{\delta}
\end{align}
for the phase and 
\begin{align}\label{eq:fixedamp}
    r_* = \sqrt{\frac{4\omega\Delta\gamma}{3}
    +\frac{2}{3}\sqrt{\delta^2-(2\omega\epsilon)^2}}
\end{align}
for the amplitude. This result suggests that systems away from resonance also
approach a chiral steady state with constant amplitude, as long as the (signed)
frequency shift is above the threshold for $r_*$ to be real: 
\begin{align} \label{eq:minshift}
    \Delta\gamma_\text{min} = -\frac{1}{2\omega}\sqrt{\delta^2 - (2\omega\epsilon)^2}
\end{align}
Numerical trajectories of the full trimer system away from resonance
[\figref{fig:4panel}(d)] agree with these predictions: the trimer attains a
steady-state amplitude predicted by \eqnref{eq:fixedamp} for modulation
frequencies above a threshold set by \eqnref{eq:minshift}.

The averaging result, \eqnref{eq:fixedamp}, states that the chiral steady state
is reached for indefinitely large positive detuning above the threshold
\eqnref{eq:minshift}, with a steady-state amplitude that grows as
$\sqrt{\Delta\gamma}$. This would appear to be at odds with the linear analysis
reported in \figref{fig:baselinespectra}(d), which shows that the Floquet
multipliers only exceed unity for a finite range of frequencies around the exact
resonance. We show in \appref{app:consistency} that the two analyses are
consistent, and the linear frequency response can be understood as a limit of
the nonlinear response.

\subsection{Reproduction of continuum trajectory in trimer model}

The numerical results in \figref{fig:4panel} and \figref{fig:reducedplot} have established that the trimer model, when initialized in the amplified traveling-wave mode singled out by the space-time symmetry, reliably evolves to the constant-amplitude chiral steady state predicted by the reduced averaged equations for wide ranges of the system parameters.
We now demonstrate that the reduced averaged equations can also explain dynamical trajectories observed in the physically realistic continuum model.
To do so, we revisit the trajectory shown in \subfigref{fig:comsol}{c}, which settled into the traveling-wave mode shortly after initialization and shared the key features of early exponential growth, amplitude saturation and ringdown, and late-time chiral steady state with the trimer trajectory.
We use these features to fix parameters in the trimer model that reproduce the behavior of the plate resonator system.

The fitting procedure, which extracts different parameters from distinct dynamical features rather than a global multi-parameter fit to the entire trajectory, is detailed in \appref{app:comsolfit}.
In brief, the linear eigenmode analysis [\subfigref{fig:comsol}{b}] yields the wave frequency $\omega$ directly;
a fit of the amplitude ringdown region to a damped harmonic oscillator trajectory decaying to the predicted fixed point then allows us to identify the damping parameter $\epsilon$ and the nonlinearity parameter $\alpha$ through the expressions \eqnref{eq:linearizationfreq} and \eqnref{eq:trimerfp}.
Finally, an exponential fit to the initial amplitude growth of the continuum trajectory provides an estimate of the growth rate $g$, which fixes the modulation strength $\delta$ via \eqnref{eq:growthrate}. 
The reduced averaged equation, \eqnref{eq:reducedeom}, is numerically integrated with the fitted parameter values to generate an amplitude trajectory for the plate resonator system.
The resulting trace [thick red curve in \subfigref{fig:comsol}{c}] closely follows the amplitude evolution from the continuum simulations, showing that the discrete trimer model can quantitatively predict the time-evolution of the parametrically amplified mode in a continuum
system.

\subsection{Basin of attraction for the chiral steady state}\label{sub:damped}

\begin{movetext2}
\end{movetext2}

\begin{movetext2}
\end{movetext2}

\begin{figure}
    \centering
    \includegraphics{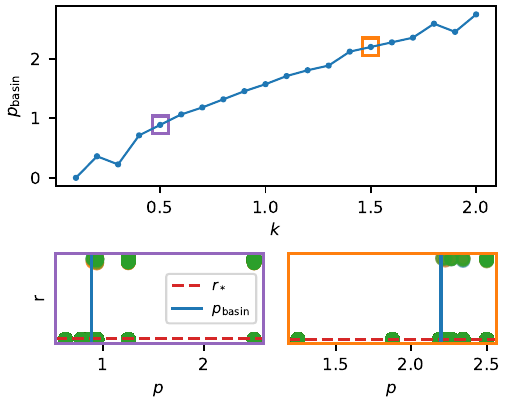}
    \caption{ Phase space radius $\pbasin$ of the basin of attraction of
   the amplitude $r_*$ in \eqnref{eq:trimerfp} as a function of the coupling
   $k$.
   Convergence to the predicted fixed-point amplitude from a set phase space distance $p$ was tested by integrating the trimer equations of motion from 300 initial conditions randomly chosen from a hypersphere centered at the fixed point with radius $p$.
   The basin of attraction radius, defined as the largest value of $p$ for which the steady-state amplitude $r$ from all numerical traces converged to the predicted value $r_*$, was estimated using a binary search along $p$  (see \appref{app:basin} for details).
   Lower panels show the convergence of the binary search to the estimated $\pbasin$ for the two points indicated by boxes in the top panel. 
   Integrations for this plot
   used the parameters $\delta = 0.6$, $\epsilon = 0.06$, and fixed the
   modulation frequency $\gamma$ at $\gamma = 2\omega$ with $\omega =
   \sqrt{1+3k}$. 
 }
    \label{fig:basin_radius}
\end{figure}

So far, we have focused on the evolution of the trimer when initialized in the
amplified traveling-wave mode, for which the chiral ansatz \eqnref{eq:ansatz}
reduces the dynamics to the two-dimensonal reduced averaged equations.
The presence of an
attractor in the reduced phase space introduces the possibility of a
corresponding attractor in the full six-dimensional phase space of the trimer,
and that initial conditions besides the amplified Floquet mode will also
converge to this attractor as long as they reside in its basin of attraction. We
now relax the restriction to the amplified mode and investigate whether the
chiral steady state can be reached from arbitrary initial conditions in the full
phase space.

To test whether the full trimer steady-state has a nonzero-volume basin
of attraction and determine its size, we numerically investigated the
convergence of the full trimer system to the chiral steady state for random
initial conditions. 
We measured the basin of attraction of the chiral steady state by
choosing a point on the attractor corresponding to the chiral state, choosing
random initial conditions at set distances from this point in the trimer's
six-dimensional phase space, and integrating for long times from those initial
conditions, recording whether or not they converged to the chiral state (see \appref{app:basin} for details). The
largest distance at which all the initial conditions converged to the chiral
state was defined to be the radius of the basin of attraction, $p_\text{basin}$,
in \figref{fig:basin_radius}. 
Since inter-oscillator
coupling is essential to single out the chiral state \eqnref{eq:trimerfp} from
other steady states with different phase relationships expected at zero or weak
coupling, we expected that the convergence to the chiral state would increase
with coupling constant $k$. (See \appref{app:perturb} for a perturbative
description of how increasing the coupling singles out the chiral state.)

\Figureref{fig:basin_radius} shows the numerically-obtained basin of attraction for a range of couplings in the
vicinity of the baseline value $k=1$, with other parameters kept constant. We
observed a non-zero radius of convergence for all except the smallest values of
coupling, confirming that the fixed point amplitude is reached from a broad
range of initial conditions. The radius increases with coupling strength in
accordance with our expectation. For the linearized system, since the chiral
mode is the only exponentially growing mode, it always dominates at long times
regardless of the initial conditions; our numerical results suggest
that the nonlinear system reliably reaches the chiral steady-state from a wide
range of initial conditions, in analogy with the amplified mode eventually
dominating in the linear system.

Our analysis in this subsection establishes a primary conclusion of this work: a
chiral constant-amplitude steady state is obtained in the trimer for a wide
range of parameters and initial conditions. 
 The fixed point position predicted from the \textit{reduced} averaged
equations, \eqnref{eq:trimerfp}, predicts the steady-state
amplitude of the \textit{full} dynamics as long as the initial conditions lie
within basins of attraction whose size is determined by the coupling strength. 
The finite-element trajectory in \subfigref{fig:comsol}{d} 
exhibits both these features, suggesting that they also apply to the continuum plate resonator system.
The trajectory deviates significantly from the space-time symmetric ansatz at early times, yet converges to the chiral steady state at late times, consistent with a finite basin of attraction for the chiral steady state. 
In addition, the amplitude of the steady state is successfully predicted from
the discrete model fixed point [\eqnref{eq:trimerfp}] by doubling the modulation
amplitude obtained via fitting from the $\delta' = 0.1$ simulation, and keeping
the other parameters constant [thick red line in \subfigref{fig:comsol}{d}].
This result illustrates the predictive power of the discrete trimer model and the reduced averaged equations for physically realistic systems, even for trajectories that initially deviate from the chiral mode.

\section{Discussion} \label{sec:discussion}

In this work, we investigated the nonlinear dynamics of a coupled nonlinear
parametric oscillator (NPO) system with a space-time symmetric parametric
modulation that generates a single unstable Floquet mode in the linearized
system. We found that the mode evolves to a nonlinear steady state that inherits
the chirality of the unstable mode, showing that nonlinearity can restrict the
growth of Floquet instabilities while maintaining desirable symmetries. The
nonlinear evolution of the unstable mode in six-dimensional phase space is
well-described by the reduced averaged equations \eqnref{eq:realreducedeom}--
equations having the same form as the averaged equations for
a \textit{single} nonlinear Mathieu oscillator-- with the effect of coupling
captured through a rescaling of parameters. This reduction yields analytical
predictions for the parameter dependence of the steady state and the time scales
dictating the approach to it. We tested these predictions against numerical
evaluation of the dynamics in the discrete trimer model, and showed that
full-wave simulations of a continuum system of mechanical plate resonators
quantitatively reproduce these features.

While our analytical approach used the standard averaging procedure for
oscillatory systems with slow amplitude evolution, the dimensional reduction to
a single oscillator was achieved by assuming an ansatz for the full nonlinear
trajectory, \eqnref{eq:ansatz}, that respects the underlying space-time
symmetry. The resulting reduced averaged equation has the same form as the
well-studied averaged equation for a single nonlinear Mathieu
oscillator~\cite{eichler2023classical}. 
Prior studies of NPO arrays ~\cite{lifshitz2003response,Bromberg2006} have
obtained similar reduced descriptions for individual parametrically-driven
modes, but the equations assumed uniform modulation phases. Our work extends
existing results by considering the case of nonuniform, spacetime-symmetric
modulation and showing how this modulation pattern allows for an algebraic
reduction of the equations of motion. 
Furthermore, we have
presented evidence that our reduced description generalizes to systems beyond
the discrete trimer model (\figref{fig:comsol}), and when the full system is
initialized in the amplified Floquet mode, predicts the entire trajectory of the
full system [\subfigref{fig:comsol}{b} and \figref{fig:threepanel}], not just
the steady-state as was the focus in Refs.~\onlinecite{lifshitz2003response} and
\onlinecite{Bromberg2006}.

We also found that many aspects of the chiral steady state persist for dynamical
trajectories that are initialized away from the space-time symmetric form. We
numerically confirmed that the chiral steady state possesses a finite basin of
attraction in the expanded six-dimensional phase space, whose radius grows with
the coupling strength, showing that the chiral state can be consistently reached
without fine-tuning of initial conditions. At the level of individual
oscillators, the steady state is stabilized by a specific phase relationship
between the motion and the underlying parametric modulation that balances
parametric gains and
losses~\cite{Goto1959,eichler2023classical,apffel2024experimental}; this
relationship is satisfied by two discrete oscillation phases separated by $\pi$,
leading to $2^3 = 8$ limit cycles, or four possible relative phase
relationships, for three decoupled NPOs. Our analysis in \appref{app:perturb}
shows that under sufficiently strong inter-oscillator coupling, the space-time
symmetric modulation at twice the traveling-wave mode frequency causes these
distinct limit cycles to collapse onto a single relative phase assignment
corresponding to the chiral steady state. 
The traveling-wave mode is locally
compatible with the stabilizing phase relationship: for each individual oscillator, the modulation and the oscillation have the phase offset required to achieve a stable amplitude at long times.

This compatibility of the parametrically amplified mode with a
constant-amplitude steady state is not guaranteed: it contrasts with the recently
reported self-oscillating solutions obtained for phase-offset NPO rings
modulated at twice the \textit{fundamental} mode frequency~\cite{Mahore2026a}.
At this modulation frequency, lower than that used in our work, a smooth phase
assignment that locally satisfies the gain-loss balance at each oscillator is
globally incompatible with the closed-ring topology. As a result, the dynamics
do not settle into a steady state, instead localizing phase defects in the
response that circulate persistently in a situation termed ``dynamical
frustration'' by the authors of Ref.~\onlinecite{Mahore2026a}. Our modulation
scheme avoids dynamical frustration by targeting the traveling-wave mode, rather
than the fundamental mode, for parametric amplification; the modulation
frequency and the space-time symmetric modulation phase assignment work in
tandem to enable a constant-amplitude steady state without phase defects. This
requires the coupling strength to be strong enough to avoid overlap between the
parametric resonance regions of the two modes; at very low values of coupling,
we observe trajectories that do not settle into a constant amplitude or phase,
which could indicate dynamical frustration in our system (\appref{app:perturb}
and \figref{fig:piover3}).

The trimer also fails to settle in a constant-amplitude steady state in the
absence of damping, instead exhibiting persistent amplitude oscillations
(\appref{app:undamped}). In these trajectories, the phase of each oscillator
varies periodically around the balance point between gain and loss,
alternatively extracting energy from the modulation source and releasing this energy to the source or the environment. The period of the amplitude oscillations differs drastically from
the system oscillation frequency, but its dependence on the system parameters
and the initial amplitude can be predicted from the reduced averaged equations
[\eqnref{eq:ellipticperiodclosed} and \eqnref{eq:ellipticperiodopen}].
Persistent amplitude oscillations were previously observed in pairs of NPOs with
dissipative coupling elements~\cite{Bello2019,calvanese2019theory}, but had a
different underlying mechanism: the modulation frequency was chosen to exchange
energy between the two coupled linear modes of the system, generating beats that
periodically bring the amplitude of each oscillator to zero. By contrast, the
amplitude oscillations we report arise from the energy exchange between the
chosen traveling-wave mode and the parametric modulation, and the amplitude
remains finite at all times, except when initialized on the separating curve
described in \appref{app:undamped} which has measure zero in the phase space.

\section{Conclusions}  \label{sec:conclusions}
We have demonstrated that a ring of coupled nonlinear parametric oscillators
with chiral phase-controlled parametric modulation consistently converges to a
constant-amplitude chiral steady state over wide ranges of parameters and
initial conditions. 
The steady
state arises from a confluence of selective amplification of one of the
traveling-wave modes through phase-controlled parametric amplification,
stabilization of the amplified mode through nonlinearity, exchange of energy
among oscillators through coupling, and loss of initial conditions through
damping. 
The consistency of our results on the discrete trimer with
simulations of a continuum mechanical system 
suggests that the stable chiral states could be attainable in micro
electromechanical systems~\cite{Mathew2016,Cha2018b,Carter2023,Mestre2025}, and in other coupled
nonlinear parametric oscillator systems including optical parametric
oscillators~\cite{McMahon2016,Inagaki2016}, nonlinear electrical
circuits~\cite{Bello2019,margiani2025three}, and Faraday
waves~\cite{apffel2024experimental}. The success of the discrete model in
quantitatively reproducing the continuum dynamics also shows that discrete
models with nonlinearity and parametric drive can serve as a design tool for
generating nonlinear parametric oscillator networks with desired responses
across platforms, similar to discrete modeling approaches for passive
metamaterials~\cite{Matlack2018a}. 

The reduction of the trimer dynamics to a single averaged equation generalizes
to rings of any odd number of oscillators with the same space-time symmetry.
Since the underlying linear dynamics harbors a chiral amplified mode for all
such rings~\cite{Melkani2024}, the dimensional reduction suggests a broader
principle: space-time symmetric modulation may organize arbitrarily large rings
to effectively single-mode nonlinear chiral dynamics, with potential
implications for nonreciprocal signal processing~\cite{Fleury2014,Nassar2020}.
We will investigate this principle in forthcoming work. More broadly, our work
highlights modulation phase control as an underutilized tool to control many-body
nonlinear steady states in coupled parametric oscillator networks.

\begin{acknowledgments}
  Research reported in this publication was supported by the National Science
  Foundation through award CMMI-2128671. We thank Benjam\'in Alem\'an, Brittany
  Carter, and Uriel Hernandez for insights on the experimental system and for
  input on parameter values for the plate resonator simulations. Initial layouts
  of the finite-element model of plate resonators were developed by Brevin
  Tuttle. We acknowledge useful discussions with Benjamin Apffel, Corentin Coulais, Arnaud Lazarus, Abhijeet
  Melkani, Jason Murphy, and Wenqian Sun. Abhijeet Melkani originated the idea
  of estimating the nonlinear time scale $\tnl$ in
  \secref{sub:avgdyntimescales}, and Wenqian Sun found the elliptic function
  expressions for the period of the conservative trimer amplitude
  \eqnref{eq:ellipticperiodclosed} and \eqnref{eq:ellipticperiodopen}.
\end{acknowledgments}

\appendix

\renewcommand{\thefigure}{\thesection\arabic{figure}}

\section{Finite-element simulation of coupled plate system}
\label{app:comsol}
The full-wave simulations of the coupled membrane resonators was implemented
using the scientific analysis software \comsolfull{}, version 6.2. We recreated
a trio of resonators by defining an equilateral triangular elastic plate with a
side length of 16.5 microns and punching out an inverted triangular hole in its
center of side length 3 microns. Normal line segments from the corners of the
inner hole to the nearest outer edges defined three quadrilateral regions. The
fundamental vibrational mode of each quadrilateral defines an oscillator degree
of freedom, coupled to its neighbors by the continuous plate material. We chose
structural and material parameters to recreate typical values for micro electromechanical 
resonators based on low-dimensional materials such as graphene, listed in
Table~\ref{tab:parameters}.

\begin{table}[b]
    \centering
    \caption{Material and structural parameters used for the plate resonator model in \comsol.}
    \label{tab:parameters}
    \begin{tabular}{ll}
        \toprule
        Parameter & Value \\
        \midrule
        Plate thickness & \qty{0.345}{nm} \\
        Poisson ratio   & \num{0.456} \\
        Young's modulus & \qty{1e12}{Pa} \\
        Density         & \qty{1.8}{g/cm^3} \\
        Base in-plane stress, $\sigma_0$    & \qty{1e7}{Pa} \\
        \bottomrule
    \end{tabular}
\end{table}

Full three-dimensional thin-plate dynamics were implemented using a Shell interface in the Structural Mechanics module in \comsol{}, which uses second-order Lagrange finite elements.
The finite-element mesh was automatically generated in \comsol{} using the `Physics-controlled' option with `Extremely fine' element size, producing a mesh with 2490 triangular elements.
Clamped boundary conditions were imposed on the edges of the outer and inner triangles.
Crucially, we used a linear elastic material for the thin-shell dynamics, but included geometric nonlinearity which couples the in-plane strains $u_{ij}$ of the midsurface with transverse deformations $h(x,y)$ via the strain component $$u_{ij} = \frac{1}{2} \partial_i h \partial_j h$$
This geometric nonlinearity provides two mechanisms that are crucial for the discrete model dynamics to be recreated: it allows the fundamental mode frequency of each resonator to be modulated dynamically by varying the in-plane tension; and it generates a stiffening nonlinearity in the dynamics of the transverse field.
These mechanisms are generic to plate resonators regardless of material, and in particular they do not rely on nonlinear elasticity at the material level.

To identify the relevant mode frequency for parametric modulation, we first carried out a pre-stressed eigenfrequency study by applying a uniform in-plane isotropic stress $\sigma_{xx} = \sigma_{yy} = \sigma_0 = \qty{1e7}{Pa}$ and computing the lowest six eigenmodes of the plate.
The lowest three eigenmodes are well-separated from the higher modes, and the frequencies of modes two and three are nearly identical [\subfigref{fig:comsol}{b}]; this mirrors the expectation from coupling the fundamental modes on each resonator as in \figref{fig:trimer}.
However, \comsol{} does not decompose the degenerate subspace into the clockwise and counterclockwise-rotating modes, but rather into  arbitrary linear combinations of them with the degeneracy lifted by the mesh.

To recreate the dynamics of the discrete model in the elastic system, we modulated the in-plane prestresses within each of the three quadrilateral domains (indexed by $j$) as
\begin{align}\label{eq:stressmod}
    \sigma_j(t) = \sigma_0\left[1+ \delta' \cos\left(2\omega_1 t + \frac{2\pi}{3}j\right)\right].
\end{align}
Dissipation was incorporated by turning on Rayleigh damping for the linear elastic material with a mass damping coefficient $\alpha_\text{dM} = 100/\omega_1$, which physically recreates a quality factor of 100 for linear oscillations at the frequency of the chiral mode.
The system is initialized with displacements from the second eigenmode (which overlaps with, but is not exactly parallel to, the parametrically amplified mode).
The \textit{generalized alpha} solver, standard for inertial dynamics in \comsol{}, was used in a time-dependent study with a relative tolerance of $10^{-5}$.
We measured the vertical displacements at the centroids of the three quadrilaterals as a proxy for the amplitude of the out-of-plane deflections of the three resonators.
For the runs shown in \figref{fig:comsol}, we generated trajectories 30 microseconds long in \comsol{}, recording the resonator displacements at time steps spaced by 0.01 microseconds.
The spacing of the internal time steps is much smaller and set adaptively by \comsol{} to meet the specified tolerance.

\section{Nondimensionalizing the equations of motion} \label{app:nondim}

In terms of dimensionful quantities, the equations of motion for the ``trimer''
model shown in \figref{fig:trimer} are
\begin{align}
    &MX_i'' + EX_i' + M\Omega^2[1+\delta\cos(\Gamma T + \beta_i)]X_i \\
    &+ AX_i^3 +K(2X_i-X_{i-1}-X_{i+1}) = 0
\end{align}
Here, $M$ is the mass of each oscillator, $E$ is the damping constant, $\Omega$
is the natural frequency of the vertical springs, $A$ is the nonlinearity
parameter, $K$ is the coupling constant, $\delta$ is the parametric drive
strength (already dimensionless), $\Gamma$ is the parametric drive frequency,
$X_i$ is the displacement of the $i$th oscillator, and $T$ is the time.
Primes denote differentiation with respect to $T$, and capital letters are all
dimensionful quantities.

If we let $t = \Omega T$ and let a dot denote differentiation with respect to
$t$, we get 
\begin{align}
    &\ddot{X}_i+\frac{E}{M\Omega}\dot{X}_i 
    + \left[1+D\cos\left(\frac{\Gamma}{\Omega}t+\beta_i\right)\right]X_i \\
    &+\frac{A}{M}X_i^3 + \frac{K}{M}(2X_i-X_{i-1}-X_{i+1}) = 0
\end{align}

This prompts the definition of the dimensionless quantities 
\begin{align}
    \epsilon = \frac{E}{M\Omega},\text{ } \gamma = \frac{\Gamma}{\Omega},
    \text{ } k = \frac{K}{M\Omega^2} 
\end{align}
and the quantity
\begin{align}\label{eq:alpha}
    \alpha = \frac{A}{M\Omega^2}
\end{align}
which has units of $1/(\text{length})^2$.

In terms of these quantities, the equations of motion have the form
\begin{align}\label{eq:trimereomwithalpha}
    &\ddot{X}_i+\epsilon \dot{X}_i+[1+\delta\cos(\gamma t + \beta_i)]X_i \\
    &+ \alpha X_i^3 + k(2X_i-X_{i-1}-X_{i+1}) = 0 \nonumber
\end{align}
While not fully dimensionless, these equations demonstrate an important feature
of oscillator equations with a cubic nonlinearity: since the nonlinearity
parameter $\alpha$ has units of $1/(\text{length})^2$, choosing a value of the
nonlinearity $\alpha$ is equivalent to choosing a length unit. This means that
if we have a solution obtained for a particular value of $\alpha$, there exists
a family of solutions that are related to the known solution by a rescaling of
the positions and velocities. It also means that if we would like to study the
effect of changing the strength of the nonlinearity on the solutions for a given
set of initial conditions, this is equivalent to varying the magnitudes of the
initial conditions while holding $\alpha$ constant.

To obtain the fully dimensionless equations \eqref{eq:trimereom} used in the
main text, we define the length scale $\Xs = 1/\sqrt{\alpha}$. We then define
dimensionless displacements $x_i$ via $\Xs x_i = X_i$ to obtain the
dimensionless equations of motion \eqref{eq:trimereom}.

\section{Numerical Methods and Procedures}\label{app:methods}
\subsection{Computing trajectories}\label{sub:computingtrajectories}
We numerically integrated the trimer equations of motion using the  
\texttt{solve\_ivp} function in SciPy~\cite{scipy}. We used the default RK4 integrator with a relative
tolerance of $10^{-6}$ and an absolute tolerance of $10^{-9}$ when computing
damped trajectories, and relative and absolute tolerances of $10^{-7}$ and
$10^{-10}$, respectively, when computing undamped trajectories.

\subsection{Determining the period of a numerical integration
(\figref{fig:integralandextracted})}\label{sub:numericalperiod}
To extract the period of the amplitude oscillations from numerical trimer
trajectories (\figref{fig:integralandextracted}, dots), we computed the upper
envelope of the trajectory, used a Butterworth filter to remove high-frequency
features, and computed the time between the adjacent local maxima of the
filtered envelope. The smallest of the time intervals between adjacent local
maxima over the trajectory was reported as the numerical period; this choice
prevented errors due to missed maxima in the peak-finding algorithm.

\subsection{Parameter sweeps in \secref{sub:damped}}\label{sub:paramsweeps}
For the integrations where the system parameters were varied (Figures
\ref{fig:4panel} and \ref{fig:reducedplot}), we varied the system parameters
$k$, $\delta$, $\epsilon$, and $\gamma$ one at a time around a set of baseline
parameters. When each parameter was varied, the others were kept at their
baseline values, with the exception that when $k$ was varied, the modulation
frequency $\gamma$ was also varied such that $\gamma = 2\omega$ unless otherwise
specified. For numerical experiments that varied one of the system parameters,
the system was initialized in the unique amplified Floquet eigenstate of the
linearized system (the system \eqref{eq:trimereom} without the cubic term),
rescaled to have a norm of 0.1, except in cases where random initial conditions
were chosen. The Floquet eigenstate depends on the parameters. 

In all trials, the steady-state amplitude was estimated from the numerical
integrations of \eqnref{eq:trimereom} by taking the maximum value of each
oscillator's position in an interval close to the end of the integration time.
For \figref{fig:4panel}, this interval was the last 200 dimensionless time
units of the 10000-unit integration window. For \figref{fig:reducedplot},
the interval was also the last 200 dimensionless time units, but the total
integration time was 2000 units because the amplitude was observed to converge
more quickly for these parameters.

\section{Derivation of the reduced averaged equation}\label{app:raederivation}

In this appendix, we derive the reduced averaged equation
\eqnref{eq:reducedeom}. We formulate general averaged equations for weakly
coupled oscillators~\cite{strogatz_nonlinear_1994,sanders_averaging_2007,guckenheimer1983nonlinear} and use them to average the equations of motion
\eqnref{eq:trimereom}. Finally, we substitute the traveling-wave ansatz
\eqnref{eq:ansatz} into the averaged equations, which reduces these three equations
into three identical copies of a single equation.

To compute the averaged equations, we first express the equations of motion in
the general form of weakly nonlinear, weakly coupled oscillators:
\begin{align}\label{eq:nonlinearoscillator}
    \ddot{x}_j + x_j + h_j(\vec{x},\dot{\vec{x}},t) = 0
\end{align}

Here, $h_j(\vec{x},\dot{\vec{x}},t)$ includes the parametric drive, damping,
nonlinearity, and coupling terms:
\begin{align}\label{eq:hj}
    h_j(\vec{x},\dot{\vec{x}},t) &= \delta\cos(\gamma t + \beta_j)x_j \\
    &+ \epsilon\dot{x}_j + x_j^3 + k(2x_j-x_{j+1}-x_{j-1}) \nonumber
\end{align}

Next, we represent the state of the system in terms of complex, time-dependent amplitudes $A_j$ that capture the ``slow'' dynamics of each oscillator relative to the fast oscillations at frequency $\omega = \sqrt{1+3k}$. Upon applying the coordinate change described in equations \eqref{eq:complexcoordspos} and \eqref{eq:complexcoordsvel} to the equations of motion, \eqnref{eq:nonlinearoscillator} yields 
\begin{align}\label{eq:exactcomplexamp}
    \dot{A}_j = \frac{1}{2\omega}ih_je^{-i\omega t} + 
    \frac{1-\omega^2}{2\omega}i(A_j+A_j^*e^{-2i\omega t}).
\end{align}
We then time-average both sides over a period $2\pi/\omega$ to obtain an evolution equation for the complex amplitudes,
\begin{align}\label{eq:genavgcomplexamp}
    \dot{A}_j = \frac{1}{2\omega}i \langle h_je^{-i\omega t}\rangle + 
    \frac{1-\omega^2}{2\omega}iA_j.
\end{align}
Here the arguments $x_j$ and $\dot{x}_j$ to the function $h_j$  are to be
expressed in terms of $A_j$ and $A_j^*$ as given by Equations \eqref{eq:complexcoordspos} and \eqref{eq:complexcoordsvel}, and angle brackets indicate time averaging over
the period $2\pi/\omega$.
When averaging, we treat the amplitudes $A_j$ as
constants and assume $\langle dA_j/dt \rangle  \approx d \langle A_j
\rangle/dt$, approximations that hold as long as $h_j$ is sufficiently
small~\cite{sanders_averaging_2007}.

We now apply \eqnref{eq:genavgcomplexamp} to the trimer. As explained by the
linear Floquet analysis (\secref{sec:linear}), the chiral modulation singles out
the counterclockwise traveling-wave mode for parametric resonance, which occurs
within a window of modulation frequencies near $2\omega$ as seen in
\subfigref{fig:baselinespectra}{d}. To keep the modulation frequency within this
resonance window, we set $\gamma = 2\omega + \Delta\gamma$, where $\Delta\gamma$
is a frequency shift with the same order of smallness as the other parameters.
Inserting \eqnref{eq:hj} into \eqnref{eq:genavgcomplexamp}, we obtain averaged
equations for the amplitude:
\begin{align}\label{eq:complexavgeoms}
    \omega \dot{A}_j &= 
    \frac{1}{4}i\delta A_j^*e^{i(\Delta \gamma t + \beta_j)} \\
    &-\frac{1}{2}\omega\epsilon A_j + \frac{3}{2}i|A_j|^2A_j \nonumber \\
    &+ \frac{1}{2}ik(2A_j-A_{j-1}-A_{j+1})+\frac{1}{2}i(1-\omega^2)A_j \nonumber
\end{align}

Up to this point, the three degrees of freedom oscillate with the same frequency
$\omega$, but with different amplitudes and phases encoded in the separate
variables $A_j$. However, the trajectories in \figref{fig:threepanel} appear to
maintain similar amplitudes and oscillation phase differences among the three
oscillators at all times.
Motivated by this observation, we substitute an ansatz with equal amplitudes
separated by $2\pi/3$ in phase:
$$A_j = Ae^{-2\pi i (j-1)/3},$$
i.e., the amplitude variables are simply offset by a phase of $2\pi/3$~\footnote{We add
an overall $(-)$ sign to the phases because the amplified mode travels in the
direction opposite the modulation wave~\cite{Melkani2024}.}. 
Under this substitution the three amplitude equations all become the same
equation, \eqnref{eq:reducedeom} in the main text, which we call the reduced averaged equation.

\section{Derivations of time scales of the nonlinear trajectory}\label{app:nlderivations}
Here we derive the initial growth rate of the amplitude of the unstable Floquet mode, and the frequency of the amplitude ringdown to the chiral steady state.

We start by specializing to the resonance case in the real reduced averaged
equations \eqnref{eq:realreducedeom}, setting $\Delta\gamma = 0$. This gives
\begin{align}\label{eq:realreducedeomresonance}
    \dot{r} &= \frac{\delta}{4\omega} r \sin(2c)-\frac{1}{2}\epsilon r \\
    \dot{c} &= \frac{\delta}{4\omega} \cos(2c) + \frac{3}{8\omega} r^2 \nonumber
\end{align}
Equations \eqref{eq:realreducedeomresonance} constitute an autonomous, planar,
nonlinear system for the averaged amplitude $r$ and phase $c$.

In the limit when the amplitude is small, the system is approximately linear and
Floquet theory applies, which predicts an exponentially-growing unstable mode. Therefore, the exponential growth can be understood from the linear limit of the dynamics, obtained by removing the $r^2$ term from
\eqnref{eq:realreducedeomresonance} which arises from the
nonlinearity in the original equations of motion:
\begin{align}
    \dot{r} &= \frac{\delta}{4\omega} r \sin(2c)-\frac{1}{2}\epsilon r \label{eq:linearizedrealreducedeom} \\
    \dot{c} &= \frac{\delta}{4\omega} \cos(2c) \label{eq:linearizedrealreducedeomphase}
\end{align}
Setting the RHS of \eqnref{eq:linearizedrealreducedeomphase} to zero yields two unique fixed points for the phase (modulo $2\pi$): $c = \pi/2 \pm \pi/4$, for which $\sin(2c) = \pm 1$. 
These fixed points correspond to the exponentially growing Floquet mode and its exponentially decaying partner.
By setting $\sin(2c)=1$ in \eqnref{eq:linearizedrealreducedeom}, we can read off the growth rate for the exponentially growing solution $r(t) \sim e^{gt}$ as
$$g = \frac{\delta}{4\omega} - \frac{\epsilon}{2},$$
which is \eqnref{eq:growthrate}.

Equations \eqref{eq:linearizedrealreducedeom} are also the result of
averaging the damped, \textit{linear} Mathieu equation
\cite{kovacic2018mathieu}: 
\begin{align}\label{eq:unaveraged}
    \ddot{x} + \left(1+\frac{\delta}{\omega}\cos 2t\right)x + \epsilon \dot{x} = 0.
\end{align}
The linear behavior of the amplified mode in our coupled system can therefore be understood using the well-known results for the Mathieu equation. 
In particular, \eqnref{eq:growthrate} matches the perturbative result for the exponential growth rate of the primary resonance of the Mathieu equation with damping~\cite{kovacic2018mathieu}.

To derive the natural frequency of the amplitude ringdown, we linearize the polar reduced averaged equations at resonance,
\eqnref{eq:realreducedeomresonance}, around the fixed point \eqnref{eq:trimerfp}.
To do so, we write the real amplitude and phase as
\begin{align}
    r &= r_* + s \\
    c &= c_* + q,
\end{align}
where $s$ and $q$ are deviations from the fixed point coordinates $r_*$ and $c_*$ obtained by setting
$\dot{r}=0$ and $\dot{c}=0$ in \eqnref{eq:realreducedeomresonance} [see
\eqnref{eq:trimerfp}].
Upon substituting these forms into \eqnref{eq:realreducedeomresonance} and keeping terms to first order in $s$ and $q$, we obtain the linearized equations of motion 
\begin{align}
    \begin{bmatrix}
        \dot{s} \\ \dot{q}
    \end{bmatrix} = 
    \begin{bmatrix}
        0 & (\delta r_*/(2\omega)) \cos(2c_*) \\
        3 r_*/(4\omega) & -\epsilon 
    \end{bmatrix}
    \begin{bmatrix}
        s \\ q
    \end{bmatrix}
\end{align}
By differentiating the equation for $q$, these can be further simplified to a
single equation:
\begin{align}\label{eq:bphase}
    \ddot{q}+\epsilon\dot{q}+\left[\left(\frac{\delta}{2\omega}\right)^2-\epsilon^2\right]q = 0
\end{align}
which is the equation for a damped harmonic oscillator.
Therefore, the ringdown to the steady amplitude shows damped oscillations with natural frequency
\begin{align}\label{eq:linearizationfreq_}
    \OmegaRD = \sqrt{(\delta/(2\omega))^2-\epsilon^2}    
\end{align}
and damping constant $\epsilon/2$; i.e. the oscillations die out over a time that scales as $~1/\epsilon$.

\section{Undamped system: persistent amplitude oscillations}\label{app:undamped}

\begin{figure}
    \centering
    \includegraphics{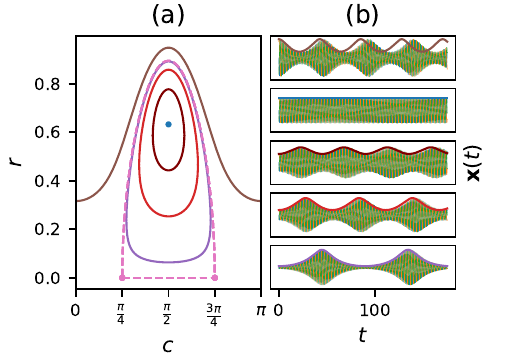}
    \caption{(a): Phase portraits of the reduced averaged equations. The initial
    conditions for the amplitude index the different curves; the phase is
    initialized at the fixed point value. (b): Numerical solutions to the full
    equations of motion compared to the amplitude found by solving the reduced
    averaged equation. The colors of the amplitudes correspond to the colors of
    the phase curves in the left figure. The vertical range shown in the plots
    at right is $-1.1\leq x_j\leq 1.1$. As the system moves deeper into the
    nonlinear regime (higher amplitudes), the averaging approximation should
    break down; this may explain the deviation between the averaged and exact
    trajectories at the top of the right figure. The parameter values are
    $\epsilon = 0$, $\delta = 0.6$, $k = 1$, $\gamma = 2\omega = 2\sqrt{1+3k} =
    4$. }
    \label{fig:undamped}
\end{figure}

When the damping $\epsilon$ approaches zero, the decay time of the ringdown oscillations to the steady state diverges; 
 i.e., the amplitude never converges to a steady value but instead exhibits persistent oscillations. 
 These amplitude oscillations represent a distinct class of nonlinear trajectories that are also predicted by the reduced averaged equations in the absence of damping. 
We now analyze these persistent oscillations and verify their existence in the full dynamics of the system.
Unless explicitly stated otherwise, we assume resonance conditions
throughout this section: $\Delta\gamma = 0$.
In addition, we
consider the phase space of the reduced averaged equations on resonance, \eqnref{eq:realreducedeomresonance} as having a
cylindrical topology with the $c$-axis wrapped around the cylinder; i.e.,
$c$-values are considered mod $\pi$, which is the period of
\eqnref{eq:realreducedeomresonance} in the $c$-direction.

\subsection{Fixed point, conserved quantity, and phase plane separation}

When $\epsilon=0$, \eqnref{eq:realreducedeomresonance} has a fixed point at 
\begin{equation} \label{eq:undampedfixedpt}
 (r_*,c_*) = \left(\sqrt{\frac{2\delta}{3}},\frac{\pi}{2}\right).  
\end{equation}
This expression is also recovered upon setting $\epsilon = 0$ in the fixed point expression for the damped system, \eqnref{eq:trimerfp}.

Away from the fixed point, the upper phase plane ($r>0$) of \eqnref{eq:realreducedeomresonance} separates into two regions in the
absence of damping: a region where the
phase curves close around the fixed point above, and a region where they do not. 
This stems from the fact that the undamped reduced averaged equations on resonance admit a
conserved quantity. Define $a = \delta/(4\omega)$ and $b = 3/(8\omega)$; then
the quantity 
\begin{align}\label{eq:conserved}
    C = \frac{br^4+2ar^2\cos(2c)}{16 b}
\end{align}
is a constant of the motion, as can be seen by taking $dC/dt$ and inserting
\eqnref{eq:realreducedeomresonance} with $\epsilon = 0$. Up to a constant
multiplier, this conserved quantity is the polar form of the rotating-frame
Hamiltonian for the averaged nonlinear Mathieu equation found in Ref.~\onlinecite{eichler2023classical}. The undamped fixed point,
\eqnref{eq:undampedfixedpt}, is a minimum of $C$ in the phase plane, which makes
it a nonlinear center~\cite{strogatz_nonlinear_1994}. We can use
\eqnref{eq:conserved} to characterize the phase curves, which are curves of
constant $C$. Defining the constant $B$ as 
\begin{equation}
    B = \frac{4b}{a}\sqrt{-C}
\end{equation}
we have from \eqnref{eq:conserved} that in the upper half-plane $r>0$,
\begin{equation}\label{eq:rfromB}
    r = \sqrt{\frac{a}{b}}\sqrt{-\cos(2c)\pm\sqrt{\cos^2(2c)-B^2}}
\end{equation}
This in turn allows us to compute $\dot{c}$ from \eqnref{eq:realreducedeomresonance}, yielding
\begin{equation}\label{eq:dcdt}
    \frac{dc}{dt} = \pm a\sqrt{\cos^2(2c)-B^2}
\end{equation}
\eqnref{eq:dcdt} shows that if $C > 0$ (so that $B$ is pure imaginary), the
phase $c$ must always be increasing [positive solution in \eqnref{eq:dcdt}] in
order to obtain a real amplitude $r$.  Therefore, $C>0$ for open curves.
Conversely, if $C < 0$ (so that $B$ is real), $c$ will both increase and
decrease over the course of one period; these correspond to the closed curves that encircle the fixed point \eqnref{eq:undampedfixedpt} (which is a minimum of $C$ with value $-a^2/b < 0$)
 Finally, there
will be a curve for which $C = 0$ that separates the open curves from the closed
curves. This curve is a heteroclinic orbit connecting two unstable fixed points
at $(r,c) = (0,\pi/4)$ and $(r,c) = (0,3\pi/4)$. From \eqnref{eq:rfromB},
its equation is 
\begin{equation}
    r = \sqrt{-\frac{2a}{b}\cos(2c)}
\end{equation}
\figref{fig:undamped}{a} shows the fixed point \eqnref{eq:undampedfixedpt} and a
selection of closed (burgundy, red, purple) and open (brown) phase curves, along
with the separating curve and the two unstable fixed points (pink) for a representative set of trimer parameters.

The conserved quantity and its associated phase curves are derived from the reduced
equations of motion. To compare these dynamics to the dynamics of the full
equations of motion, we generated initial conditions for all three oscillators
for a given initial $(r,c)$ pair using the ansatz \eqnref{eq:ansatz} and
numerically evaluated the resulting trajectories from \eqnref{eq:trimereom}. The
full trajectories follow the expectation from the averaged trajectories
[\subfigref{fig:undamped}{b}]: the high-frequency oscillations of the three
trimers maintain $2\pi/3$ phase offsets while experiencing slow amplitude
variations that closely track the predicted $r(t)$, with some deviations for the trajectory corresponding to the open phase curve.

\subsection{Period of amplitude oscillations}

\begin{figure}
    \centering
    \includegraphics[width=0.47\textwidth]{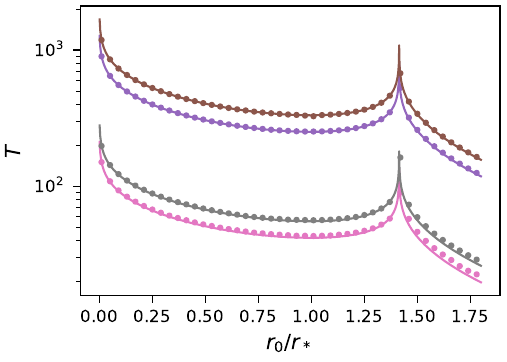}
    \caption{Comparison of the integral expression for the trimer period
    [Equations \eqref{eq:ellipticperiodclosed} and
    \eqref{eq:ellipticperiodopen}, solid lines] with periods extracted from
    numerical integration of the full equations of motion \eqnref{eq:trimereom} initialized at different amplitudes $r_0$.
    The initial phase value $c_0$
    is $\pi/2$. We use the ansatz \eqnref{eq:ansatz}, the complex-to-polar
    coordinate change \eqnref{eq:rctocomplexamp}, and the complex amplitude
    definitions \eqnref{eq:complexcoordspos} and \eqnref{eq:complexcoordsvel} to
    translate from $r_0$ and $c_0$ into initial conditions for the trimer. Solid
    curves show the period predicted by the integral expression and dots show
    the numerically-computed period. Colors distinguish different system parameters: $k = 1, \delta = 0.1$
    (purple); $k = 2, \delta = 0.1$ (brown); $k=1,\delta=0.6$ (pink);
    $k=2,\delta=0.6$ (gray). All traces have zero damping (i.e., $\epsilon =
    0$). See Appendix \ref{app:methods}, \subsectionref{sub:numericalperiod} for
    a description of the numerical period extraction method.
  }
    \label{fig:integralandextracted}
\end{figure}

The evolution equation of the phase along the curves of constant $C$, \eqnref{eq:dcdt}, can be rearranged to derive analytical formulae for the periods of both open and closed phase curves.
We rearrange \eqnref{eq:dcdt} to obtain the differential relation
\begin{equation}\label{eq:dtundamped}
    dt = \frac{dc}{\pm a\sqrt{\cos^2(2c)-B^2}}
\end{equation}
which we can integrate to find the period.

For the closed curves, we must identify the turning points of the phase curves
(where $\dot{c} = 0$) in order to integrate \eqnref{eq:dtundamped}. These are at
\begin{align}\label{eq:turningpts}
    c_\pm = \frac{\pi}{2} \pm \frac{1}{2}\cos^{-1}(B)
\end{align}
Integrating both sides of \eqnref{eq:dtundamped} with these as the limits and
multiplying by 2 (to account for traversing the parts of the phase curve both
above and below the turning points) yields
\begin{align}\label{eq:periodclosed}
    T = \frac{2}{a}\int_{B}^{1}\frac{du}{\sqrt{(u^2-B^2)(1-u^2)}}.
\end{align}
By performing a variable substitution
\begin{align}
    v = \sqrt{\frac{u^2-B^2}{1-B^2}},
\end{align}
\eqnref{eq:periodclosed} can be expressed in terms of the
of the complete elliptic integral of the first kind~\cite{abramowitz1948handbook},
\begin{align}\label{eq:ellipticK}
    K(m) = \int_0^1\frac{dv}{\sqrt{(1-v^2)(1-mv^2)}}
\end{align}
to obtain the final closed form for the period of the phase curves with $C < 0$:
\begin{align}\label{eq:ellipticperiodclosed}
    T_\text{closed} = \frac{8\omega}{\delta B}\ellipticK\left(\frac{B^2-1}{B^2}\right)
\end{align}

For the open curves with $C > 0$, we define a new real constant
\begin{align}\label{eq:scriptBdef}
    \mathcal{B} = \frac{4b}{a}\sqrt{C} = -iB
\end{align}
and with this definition \eqref{eq:dtundamped} becomes
\begin{align}
    dt = \frac{dc}{a\sqrt{\cos^2(2c) + \mathcal{B}^2}}
\end{align}
Since $\dot{c} > 0$ everywhere for the open curves, there are no turning points and
the curve traverses the entire circumference of the cylindrical phase space,
which has length $\pi$. Therefore, we set the limits of integration to $0$ and
$\pi$ and integrate \eqnref{eq:dtundamped}, yielding
\begin{align}\label{eq:periodopen}
    T = \frac{2}{a}\int_{0}^{1}\frac{du}{\sqrt{(u^2+\mathcal{B}^2)(1-u^2)}}
\end{align}
This can be rewritten  using \eqnref{eq:ellipticK} as
\begin{align}\label{eq:ellipticperiodopen}
    T_\text{open} = \frac{8\omega}{\delta\mathcal{B}}\ellipticK\left(-\frac{1}{\mathcal{B}^2}\right)
\end{align}

Note that in the limit
as $B\rightarrow 1$, which corresponds to approaching the fixed point, the
closed curve period \eqnref{eq:ellipticperiodclosed} converges to the expression for the ringdown frequency calculated from the linearization around the steady state, \eqnref{eq:linearizationfreq}, with $\epsilon = 0$. The formulae \eqref{eq:ellipticperiodclosed} and \eqref{eq:ellipticperiodopen} show the
dependence of the period of the reduced averaged equations on the coupling
through the traveling-wave frequency $\omega = \sqrt{1+3k}$. The period grows as
the coupling grows, consistent with the reduction in the ringdown frequency with
increased coupling shown in \subsectionref{sub:avgdyntimescales}. In
contrast, the fixed point formula shows that the location of the fixed point
does not depend on the coupling when the system is undamped. 

In \figref{fig:integralandextracted}, we compare the elliptic integral expressions
above with numerical computations of the period of the amplitude oscillations extracted from the full
trimer equations of motion. The analytical expressions agree with the numerically extracted values across a wide range of initial conditions ($r_0$ values) and for different system parameters (colors). 
The divergence in the oscillation period as the heteroclinic orbit at $r_0/r_* = \sqrt{2}$ is approached is apparent in the curves.

\section{Chirality of the system at various parameter values and initial conditions} \label{app:chirality}

\begin{figure}
    \centering
    \includegraphics{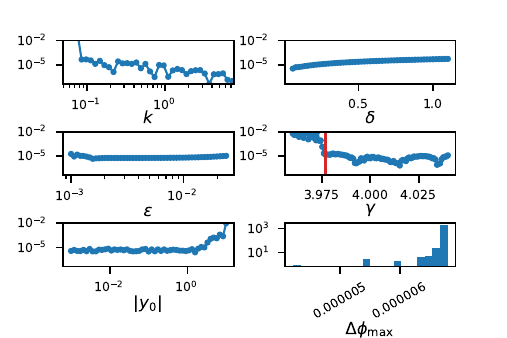} 
    \caption{Deviation from perfect chirality $\Dphimax$ as different system parameters are varied.
    Unless varied, parameters are set to $\epsilon
        = 0.01$, $\delta = 0.1$, $k = 1$, and $\gamma = 2\omega = 2\sqrt{1+3k} =
        4$, and the amplified Floquet eigenstate scaled to magnitude 0.1 provides
        the baseline initial conditions. 
        For each plot except the bottom right, the vertical axis shows the
        value of $\Dphimax$, the maximum deviation from perfect chirality, as a
        function of the variable on the horizontal axis. 
        In the top four plots, the parameter
        named in the horizontal axis label is varied. In the lower left plot, the magnitude of the initial condition vector is varied with its direction in phase space fixed to the baseline.
        The bottom-right plot is a histogram of $\Dphimax$ values obtained from 2000 runs with initial condition vector chosen randomly from the surface of a sphere in phase space with magnitude fixed at 0.1.
        For the
        varied $k$, varied $\epsilon$, and varied $y_0$ direction plots, the
        integration interval was [0,10000]; for the varied $\delta$, $\gamma$,
        and $|y_0|$ plots the interval was [0,2000]. The last 1000 points of the
        time traces were sampled to compute the chirality. In the varied
        $\gamma$ plot, parametric amplification only occurs for $\gamma$ values
        greater than the value at the red line.}
    \label{fig:chirality}
\end{figure}

Chirality--- a breaking of mirror
symmetry in the trajectories---presents in the trimer system as individual oscillations that peak and fall in a
definite order and with definite relative phases.
To assess chirality from numerical trajectories, we define a chirality
parameter
\begin{align}
    \mathcal{C}(\phi_1,\phi_2,\phi_3) = \left|\left<\sum_{j=1}^3e^{i\phi_j}\right>\right|
\end{align}
where $\phi_j$ is the phase of the complex amplitude $A_j$ of the $j$th
oscillator as defined in \eqnref{eq:complexcoordspos} and
\eqnref{eq:complexcoordsvel}. Brackets denote averaging over a time interval that
occurs after the trajectories have approximately reached a steady amplitude.
With this definition, the chirality is zero if the three phases are separated by
$2\pi/3$. Note that a chirality parameter \textit{greater than zero} corresponds
to a motion in which the oscillator phases are \textit{not} evenly spaced.

The physical significance of this parameter is not obvious for nonzero values,
so we further define a geometric parameter $\Dphimax$. If three phases
$\phi_1$, $\phi_2$, and $\phi_3$ make the sum $|\sum_{j=1}^3e^{i\phi_j}|$ equal
to $M$, and we are given that $\phi_2$ and $\phi_3$ are $2\pi/3$ and $4\pi/3$,
respectively, the value of $\phi_1$ is given by
\begin{align}
    \cos\phi_1 = \frac{2-M^2}{2}
\end{align}
Motivated by this observation, we define $\Dphimax$ via 
\begin{align}
    \cos\Dphimax = \frac{2-C^2}{2}
\end{align}
$\Dphimax$ is a proxy for the largest possible deviation from
evenly-spaced phases. Plots of $\Dphimax$ for various values of the system
parameters $k$, $\epsilon$, $\delta$, $\gamma$, as well as different values of
the initial conditions, are shown in \figref{fig:chirality}. For each
plot, the maximum $\Dphimax$ is less than approximately 0.01, and for most
parameter values shown $\Dphimax < 10^{-4}$, showing that the phases remain very
close to evenly spaced at long times.

\section{Linear frequency response as a limit of the nonlinear response}\label{app:consistency} 

\begin{figure}
    \centering
    \includegraphics{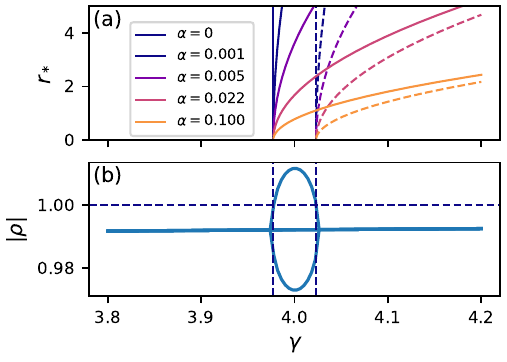}
    \caption{In the limit of zero nonlinearity, the stable and unstable
    steady-state amplitude curves collapse onto the boundaries of the Floquet
    amplification region. (a): The stable steady-state amplitudes $r_*$ (solid
    lines) and unstable steady-state amplitudes $\tilde{r}_*$ (dashed lines) at
    varied nonlinearity $\alpha$. (b): The Floquet multipliers of the trimer.
    Vertical dashed lines are the bifurcation frequencies $\gamma_-$ and
    $\gamma_+$ where the stable and unstable fixed points emerge, respectively.
    The horizontal dashed line is at $|\rho| = 1$. Note that the vertical lines intersect
    $|\rho|=1$ exactly where the Floquet multipliers equal 1 in magnitude,
    indicating that the bifurcation frequencies obtained via averaging mark the
    boundaries of the amplification window obtained via Floquet theory. }
    \label{fig:responselimit}
\end{figure}

As described in \secref{sec:linear}, Floquet theory predicts that the linearized
trimer will experience exponentially amplifying oscillations that continue to
grow for all time as long as the modulation frequency $\gamma$ lies within a
particular range. This range is the interval of frequencies where the Floquet
multipliers exceed 1 in magnitude and is shown in
\figref{fig:baselinespectra}(d) for a particular set of parameter values. But in
\figref{fig:4panel}(d), the steady-state amplitude continues to grow even when
$\gamma$ leaves the amplified region shown in \figref{fig:baselinespectra}(d).
This would appear to be inconsistent with Floquet theory, since in the nonlinear
model amplified states are accessible even when the modulation frequency exceeds
the maximum amplified frequency. 
In this
Appendix we show that the response of the linear system can be understood as a
zero-nonlinearity limit of the nonlinear response.

We first restore the dimensionful nonlinearity parameter $\alpha$ in the
equations of motion (see \appref{app:nondim}). This parameter represents the
strength of the nonlinearity, with larger $\alpha$ corresponding to a stronger
nonlinearity. With $\alpha$ restored, the steady-state amplitude in
\eqnref{eq:fixedamp} becomes 
\begin{align}\label{eq:fixedampalpha}
    r_* = \sqrt{\frac{4\omega\Delta\gamma}{3\alpha}
    +\frac{2}{3\alpha}\sqrt{\delta^2-(2\omega\epsilon)^2}}
\end{align}

Provided $\Delta\gamma$ is sufficiently large, the same expression with a minus
sign in front of the second term also produces a real root, which we label
$\tilde{r}_*$:
\begin{align}\label{eq:unstablefixedampalpha}
    \tilde{r}_* = \sqrt{\frac{4\omega\Delta\gamma}{3\alpha}
    -\frac{2}{3\alpha}\sqrt{\delta^2-(2\omega\epsilon)^2}}
\end{align}

The amplitude $r_*$ corresponds to a stable constant-amplitude state, and
$\tilde{r}_*$ to an unstable one~\cite{eichler2023classical}.

Let
\begin{align}
    \Delta\gamma_* = \frac{1}{2\omega}\sqrt{\delta^2-(2\omega\epsilon)^2}.
\end{align}
From \eqnref{eq:fixedampalpha}, the minimum frequency at which the stable fixed
amplitude $r_*$ exists is $\gamma_- = 2\omega - \Delta\gamma_*$. Similarly, from
\eqnref{eq:unstablefixedampalpha} the minimum frequency at which the unstable
amplitude $\tilde{r}_*$ exists is $\gamma_+ = 2\omega + \Delta\gamma_*$.
Applying Floquet theory to the Mathieu equation \eqnref{eq:unaveraged} says that
the steady-state amplitude of the linear trimer is infinity for $\gamma_- <
\gamma < \gamma_+$ and zero otherwise~\cite{landau1960mechanics}. In short, these
bifurcation frequencies of the nonlinear trimer correspond to the limits of the
amplification window of the linear trimer. In the limit as the nonlinearity
$\alpha$ goes to zero, we would therefore expect the response function of the
trimer to go to two vertical lines at $\gamma_+$ and $\gamma_-$. From equations
\eqref{eq:fixedampalpha} and \eqref{eq:unstablefixedampalpha}, the stable and
unstable steady-state amplitudes do converge to these vertical lines in this
limit, in the sense that for any $r > 0$ and $\varepsilon > 0$ we can choose an
$\alpha > 0$ such that the maximum frequency deviation of the $r_*$ and
$\tilde{r}_*$ curves from $2\omega \pm \Delta\gamma_*$ is less than
$\varepsilon$. This is illustrated in \figref{fig:responselimit}. While this
does not explain the presence of amplification in the region outside
$[\gamma_-,\gamma_+]$, it shows consistency between the linear and nonlinear
frequency responses in the appropriate limit.

\section{Fitting procedure for Figure \ref{fig:comsol}}\label{app:comsolfit}

Below we describe the procedure used to produce the
trajectory of the reduced averaged equations (the red curve) for the simulations reported in \subfigref{fig:comsol}{a}. We obtain this curve by integrating \eqnref{eq:reducedeom} with parameters extracted from the \comsol{} traces.
We also describe how the fit is used to predict the final amplitude of the chiral motion in \subfigref{fig:comsol}{b}.
In this section, we restore the nonlinearity parameter $\alpha$ to the trimer equations of motion
as described in Appendix \ref{app:nondim},
\eqnref{eq:trimereomwithalpha}. 

The procedure in this section finds values for values of $\delta$, $\omega$,
$\epsilon$, and $\alpha$ to be used in \eqnref{eq:reducedeom}. We assume
resonance conditions, as are used to generate the FEM traces: $\gamma =
2\omega$.

\subsection{Extract a single amplitude trace}
The finite element method (FEM) traces are produced by recording the vertical
displacement of a particular point on each plate. We first process these three
traces to produce a single amplitude trace. The procedure is as follows:
\begin{enumerate}
    \item Numerically compute the local maxima of each continuum trace to obtain
    three amplitude traces.
    \item Aggregate the three numerical amplitude traces to create a single trace, ordered by
     time.
    \item Generate a piecewise-linear interpolant of the aggregated traces to
     establish a constant time increment.
    \item Filter the interpolant with a Savitsky-Golay filter. We use the
     \texttt{savgol\_filter} function of signal module within the Python library SciPy~\cite{scipy}.
     This produces the FEM amplitude trace. We use a second-order filter with a
     window width corresponding to ten times the period $2\pi/\omega_0$ of the
     trimer fundamental mode. For \figref{fig:comsol}, $\omega_0 = \qty{52.607}{MHz}$, and
     the dimensionless frequency of the chiral mode is 
     \begin{align}\label{eq:omegafromcomsol}
     \omega = \omega_1/\omega_0 = 1.04297
     \end{align}
\end{enumerate}

\subsection{Rescale time and length units}\label{sub:rescale}

Next, we rescale the time and amplitude coordinates of the FEM amplitude traces
to make them dimensionless. The frequency scale for the FEM trace is the
frequency $\omega_0$ of the fundamental mode, which can be obtained directly as
an output from \comsol. We rescale the time points from the FEM trace to be
dimensionless by multiplying each time point by $\omega_0$. We choose to rescale
the length units to be in units of the fixed point length $R_*$; this is
achieved by numerically computing the value of the fixed point amplitude and
dividing all lengths by this value. We compute the value of the fixed point by
finding the average value of the FEM amplitude trace over a time interval where
it has converged; i.e., the oscillations about the fixed point have an amplitude
that is a small fraction of the fixed point amplitude. For 
\subfigref{fig:comsol}{c}, we use the interval from $t = \qty{27}{\micro\second}$ until the end of the
trace as the converged interval. The computed fixed point amplitude is $R_* = \qty{2.56824}{\nano\meter}$.

\subsection{Perform fits in different dynamic regions}

We use information from the different dynamical regions to fit the dimensionless parameters of the discrete model.
We first process the FEM amplitude trace in the ringdown region---the region
where the trace amplitude undergoes damped oscillations about the fixed point---by subtracting out the fixed point value, so that the amplitude oscillates about
0. We then fit a decaying oscillation of the form 
\begin{align}\label{eq:decayingexp}
    r(t) = Ae^{-\lambda t}\cos(\OmegaD t)    
\end{align}
to the data in this region. For \figref{fig:comsol}, we use the interval
$15\mu s \leq t \leq 25 \mu s$ as the ringdown region. The fit parameters are
$\lambda$ (the damping constant) and $\OmegaD$ (the damped frequency). We
numerically extract the amplitude $A$ by computing the maximum value of the
amplitude of the filtered trace. 

Now we use the fixed point formula \eqnref{eq:trimerfp} to establish
relationships between the fit parameters and the parameters of the trimer. We
restore the nonlinearity parameter $\alpha$ to \eqnref{eq:trimerfp} to obtain 
\begin{align}\label{eq:trimerfpwithalpha}
    r_* = \sqrt{\frac{2}{3\alpha}}\left[\delta^2-(2\omega\epsilon)^2\right]^{1/4}
\end{align}
We then have that the fixed point $r_*$ is
\begin{align}
    r_* = \sqrt{\frac{4\omega\OmegaRD}{3\alpha}}
\end{align}
where $\OmegaRD$ is the ringdown frequency from
\eqnref{eq:linearizationfreq}, $\omega$ is the ratio of the traveling wave
frequency to the fundamental frequency, and $\alpha$ is the nonlinearity
parameter. 
Recall that after rescaling as described in \subsectionref{sub:rescale}, $r_* =
1$, so $\alpha$ is dimensionless as it appears in this section. 

\eqnref{eq:decayingexp} is a solution to a damped
harmonic oscillator equation of the form 
\begin{align}
    \ddot{b} + 2\lambda \dot{b} + \OmegaRD^2 b = 0
\end{align}
with
\begin{align} \label{eq:omegaD}
    \OmegaD^2 = \OmegaRD^2 - \lambda^2
\end{align}
Comparing to \eqnref{eq:bphase} and \eqnref{eq:linearizationfreq}, we see that
\begin{align}
    \lambda = \epsilon/2,
\end{align}
i.e., the damping factor is obtained as twice the fit parameter $\lambda$.

To extract the nonlinearity parameter $\alpha$, we
first solve \eqnref{eq:omegaD} for $\OmegaRD$:
\begin{align}
   \OmegaRD = \sqrt{\OmegaD^2+\lambda^2}, 
\end{align}
then use the result to isolate $\alpha$:
\begin{align}
   \alpha = \frac{4\omega\OmegaRD}{3r_*^2}. 
\end{align}

The only remaining undetermined parameter is the modulation strength $\delta$.
Fit a function of the form 
\begin{align}
    r = r_0 e^{\kappa t}
\end{align}
to the growth region of the motion---the region where the amplitude is
exponentially growing. We use $5 \mu s \leq t \leq 8 \mu s$ as the growth
region for the data in \subfigref{fig:comsol}{c}. 

After fitting an exponential to the FEM amplitude in this region, we have 
\begin{align}
   \kappa = \frac{1}{2}\left(\frac{\delta}{2\omega} - \epsilon\right) 
\end{align}
which gives 
\begin{align}
    \delta = 2\omega(2\kappa+\epsilon)
\end{align}
Since $\epsilon$ is known from the previous fit, the parameters of the discrete
trimer model are now fully determined. For the fit in \figref{fig:comsol},
the resulting parameters are
\begin{align}\label{eq:comsolfitparams}
    \alpha = 0.05874 \\
    \delta = 0.09677 \nonumber\\
    \epsilon = 0.01318 \nonumber
\end{align}

\subsection{Integrate the amplitude equation}
To obtain the time-evolution of the chiral mode amplitude [red curve in \subfigref{fig:comsol}{c}], we numerically integrate \eqnref{eq:reducedeom}.
But first, we need to translate the FEM system initial conditions into initial
conditions for the reduced averaged equations. Note that since the ansatz used
to obtain the reduced averaged equations from the averaged equations is 
\begin{align}
    A_j = A e^{-2\pi i (j-1)/3},
\end{align}
the reduced averaged amplitude $A$ corresponds to the amplitude $A_1$ of the
full averaged equations. $A_1$ is distinguished from the other amplitudes by
having no phase offset in its parametric modulation term. This means we can use
the position $x_\text{FEM,1}$ and velocity $\dot{x}_\text{FEM,1}$ of the
corresponding FEM oscillator [i.e., the one with no phase offset in its modulation
term; see \eqnref{eq:membranetension}] to determine initial conditions for $A$,
via the coordinate transformation
\begin{align}
    A(0) = A_1(0) = \frac{1}{2}\left[x_\text{FEM,1}(0) - i\dot{x}_\text{FEM,1}(0)/\omega\right]
\end{align}
which we obtain from \eqnref{eq:complexcoordspos} and
\eqnref{eq:complexcoordsvel}. 

Once we have the correct initial conditions, we integrate
\eqnref{eq:reducedeom}, obtaining the complex amplitude $A(t)$. Recalling
\eqnref{eq:rctocomplexamp}, the real
amplitude is 
\begin{align}
    r(t) = 2|A(t)|.
\end{align}

As a final step before plotting, we restore the length and time units by
multiplying each dimensionless length by $R_*$ from \subsectionref{sub:rescale}
and each dimensionless time by $1/\omega_0$.

\subsection{Steady-state amplitude for $\delta'=0.2$}
Once the parameters are found, we can estimate the effect of changes in the
parameters on changes in the FEM steady-state amplitude. 

Let $\Rfp{0}$ be the numerically-computed steady-state amplitude of the FEM
trace for a set of FEM simulation parameters
$(\alpha_0',\delta_0',\epsilon_0',\omega_0')$. Here $\delta_0'$ corresponds to
the $\delta'$ in \eqnref{eq:stressmod}, $\omega_0'$ corresponds to $\omega_1$ in
\eqnref{eq:stressmod}, and $\alpha'$ and $\epsilon'$ are, respectively, the
geometric nonlinearity and the damping as implemented in the FEM simulations
(see \secref{sec:comsol} and \appref{app:comsol} for details).
$\Rfp{0}$ is as computed in \subsectionref{sub:rescale} of this appendix.
Similarly, let the unprimed variables $(\alpha_0,\delta_0,\epsilon_0,\omega_0)$
be the corresponding parameters for the discrete trimer model found via the
fitting methods in this section, and let $\rfp{0}$ be the discrete trimer fixed
point amplitude with these parameters as determined by
\eqnref{eq:trimerfpwithalpha}.

We would like to predict the (dimensionful) steady-state amplitude $\Rfp{1}$ of a new
FEM simulation with new parameters
$(\alpha_1',\delta_1',\epsilon_1',\omega_1')$. We can do so by assuming that the
ratio of the new FEM amplitude to the old FEM amplitude will equal the ratio of
the corresponding new and old trimer amplitudes:
\begin{align}\label{eq:newFEMfp}
    \Rfp{1} = \frac{\rfp{1}}{\rfp{0}}\Rfp{0} =\frac{r_*(\params{1})}{r_*(\params{0})}\Rfp{0}
\end{align}
In order to use this formula, we need to determine the new parameters
$\params{1}$ from the old trimer parameters and the FEM parameters. To do so, we
scale the trimer parameters using the ratio
of the new and old FEM parameters: 
$\alpha_1 = (\alpha_1'/\alpha_0')\alpha_0,$
and similarly for $\delta_1$, $\epsilon_1$, and $\omega_1$.
We substitute the scaled trimer parameters into \eqnref{eq:newFEMfp} to obtain the estimate for the new FEM fixed point
amplitude. 

To generate the amplitude estimate in \figref{fig:comsol}(d), we used the
parameters extra \eqnref{eq:comsolfitparams} and the value of $\omega$ from
\eqnref{eq:omegafromcomsol} for the unprimed, 0-subscript parameters. The FEM
parameters used in \figref{fig:comsol}(d) (ie, the primed, 1-subscript
parameters) were the same as those used in \figref{fig:comsol}(c) (the primed,
zero-subscript parameters), except that $\delta'$ was increased from 0.1 to 0.2;
ie, $\delta_1' = 0.2$, $\delta_0'=0.1$.
The resulting amplitude from \eqnref{eq:newFEMfp} is $\Rfp{1}=\qty{3.6904}{\nano\meter}$, which is indicated as the thick horizontal line in \subfigref{fig:comsol}{d}.
The predicted amplitude is within 0.13\% of the actual steady-state amplitude (\qty{3.6856}{\nano\meter}) measured from the dynamical traces.

\section{Numerical estimation of the radius of the basin of attraction}\label{app:basin}

\begin{figure}
    \centering
    \includegraphics{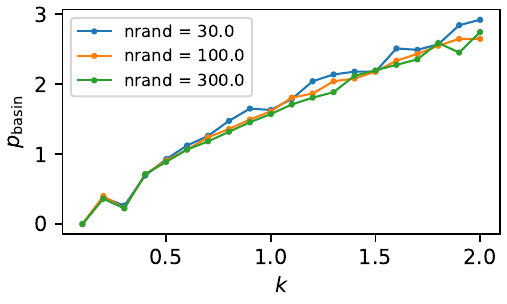}
    \caption{Convergence study for the binary search algorithm used to determine
    the basin radius $\pbasin$. The three curves show the results of the binary
    searches at each value of the coupling $k$ for 30, 100, and 300 random
    initial conditions. }
    \label{fig:basin_convergence}
\end{figure}

In this section we describe the methods used to find the phase space radius
$\pbasin$ shown in \figref{fig:basin_radius}. In summary, we found the basin
radius by solving the system with initial conditions sampled randomly from the
6-dimensional phase space of the full dynamics, choosing these initial
conditions at increasing Euclidean distances away from a point on the manifold
corresponding to the the fixed point of the reduced dynamics. We defined
$p_\text{basin}$ as the largest initialization distance from this point at which
all sampled trajectories converged to oscillatory steady states with the
predicted amplitude $r_*$.
\subsection{Background: the trimer fixed point in 6D phase space}
We first describe the constant-amplitude state corresponding to the fixed point
\eqref{eq:trimerfp} in the full six-dimensional phase space of the unaveraged
system \eqref{eq:trimereom}. The averaging procedure described in section
\ref{sec:nonlinear} captures this state as single point in a 2-dimensional phase
space whose dimensions are the ``slow'' variables $r$ and $c$ (respectively, the
amplitude of the oscillations and the phase offset). But in terms of the
original variables $x_j$ and $v_j = \dot{x}_j$, the steady state represented by
the fixed point \eqref{eq:trimerfp} is a closed curve in 6D phase space. Using
\eqnref{eq:rctocomplexamp}, \eqnref{eq:ansatz}, and the coordinate changes
\eqref{eq:complexcoordspos} and \eqref{eq:complexcoordsvel}, we can write the
positions and velocities corresponding to the ansatz \eqref{eq:ansatz} at the
fixed point:
\begin{align}
    x_j &= r_*\cos\left(\omega t + c_* - \frac{2\pi(j-1)}{3}\right)\label{eq:xvattractorx} \\
    v_j &= -\omega r_*\sin\left(\omega t + c_* -\frac{2\pi(j-1)}{3}\right) \label{eq:xvattractorv}
\end{align}
The coordinates and velocities are periodic functions of time, so the state
\eqref{eq:trimerfp}, which is a fixed point in $r$-$c$ space, is a closed loop
in $x$-$v$ phase space.

We are interested in the set of points that converge to this loop-shaped
attractor under the phase flow of the equations of motion \eqnref{eq:trimereom}.
In particular, we would like some indication that this is an open set, and how
large it is for different values of the parameters, especially the coupling $k$.
While we will not check explicitly for convergence to this phase space
loop, we will check for convergence of the trajectory \textit{amplitudes} to the
amplitude $r_*$ given in \eqnref{eq:trimerfp}, which is common to all three
oscillators when they reach the loop. In this work, we take the convergence of
trajectories to trajectories having this \textit{amplitude} as a reasonable
proxy for convergence to the loop attractor predicted by averaging.

\subsection{Defining the radius of the basin of attraction}
We define the \textit{basin of attraction of the fixed point amplitude} as the
set of all points $(x_1,x_2,x_3,v_1,v_2,v_3)$ that flow to oscillatory steady
states having the amplitude $r_*$ given in \eqnref{eq:trimerfp}. We will abuse
notation and refer to this set as the ``basin of attraction," although strictly
speaking it is not the basin of attraction of the averaging loop attractor
because it makes no reference to the oscillation phases. The basin of attraction
is a subset of $\Rn{6}$, and could have a complicated shape. As a simple
indicator of the size of the basin, we define the \textit{radius of the basin},
$\pbasin$, as the minimum distance from a particular point $\vec{y}_* \in
\Rn{6}$ on the loop attractor to the edge of the basin; we write $\vec{y}_*$ as
$y_* = (\vec{x}_*,\vec{v}_*)$ with $\vec{x}_*,\vec{v}_* \in \Rn{3}$ given by 
\begin{align}
    x_{*j} &= r_*\cos(c_* + 2\pi(j-1)/3), \\
    v_{*j} &= -\omega r_* \sin(c_* + 2\pi(j-1)/3) \\
    (j &= 1,2,3) \nonumber
\end{align}
These are the values of $x_j$ and $v_j$ from \eqnref{eq:xvattractorx} and
\eqnref{eq:xvattractorv} evaluated at $t=0$.

\subsection{Finding the basin radius as a function of $k$}
We begin by defining the \textit{indicator function} $f_I(p)$ to be 0 if any
point at distance $p$ from $y_*$ converges to a state with the fixed-point
amplitude $r_*$, and 1 otherwise. In other words, the indicator function is 0 if
the phase-space sphere of radius $p$ is entirely within the basin of attraction,
and 1 if any portion of it lies outside the basin. Note that the indicator
function is a step function with the discontinuity at the basin radius, since it
evaluates to $0$ for any $p<\pbasin$ and to 1 for $p>\pbasin$.

We cannot compute the phase flow of every point in the sphere centered at $y_*$,
so to implement $f_I$ in practice we compute the trajectories of 300
randomly-chosen initial conditions that lie in 5D hypersurface. We chose
the value 300 by conducting a convergence study with 30, 100, and 300 points,
the results of which are shown in \figref{fig:basin_convergence}. We chose each
point by 
generating a vector of six
coordinates drawn from the standard normal distribution and then normalizing the
vector to have the desired length. This procedure gives each point of the 5D
sphere the same probability of being selected because the joint probability
density for selecting each vector in this way depends only on the radius of the
sphere: $e^{-(y_1-y_{*1})^2}\cdots e^{-(y_6-y_{*6})^2} =
e^{-|\vec{y}-\vec{y}_*|^2} = e^{-p^2}$. 

We integrated the equations of motion as described in \appref{app:methods} with the trimer
parameters given in the caption of \figref{fig:basin_radius}, over an
integration time of 2000 dimensionless time units.  As
throughout this work, we compute the amplitudes of each oscillator as the
maximum displacement attained by that oscillator over a time interval near the
end of the integration window; for these trajectories, we used the last 40
dimensionless time units.
A state was judged to have converged to $r_*$ if its amplitude lay within 0.01 dimensionless distance units of $r_*$.

We use a binary search procedure to efficiently find $\pbasin$ using the
indicator function. We first set a high upper limit $L$ on the value of the
basin radius by evaluating the indicator function over a coarse, equally-spaced
range of values of $p$ beginning at zero; for the searches described here, we
chose $L=5$. We also decide on the precision $\epsilon_p$ to which we wish to
know $\pbasin$; in these searches we used $\epsilon_p = 0.01$. To complete the
search we evaluate $f_I(p)$ $n_\text{eval}$ times, on values $\{p_1 =
L/2,p_2,\cdots,p_{n_\text{eval}}\}$, where $n_\text{eval}$ is the smallest
integer such that $L/2^{(n_\text{eval}+1)} \leq \epsilon_p$. We then find the
values of $p_n$ iteratively as follows:
\begin{align}
    p_1 &= L/2 \nonumber \\
    p_{n+1} &= 
    \begin{cases}
        p_n + L/2^{n+1}, & f_I(p_n) = 0 \\
        p_n - L/2^{n+1}, & f_I(p_n) = 1
    \end{cases}
\end{align}
The value $p_{n_\text{eval}+1}$, evaluated for $k = \{0.1,0.2,\cdots,2\}$ is reported as $\pbasin$ in the upper panel of \figref{fig:basin_radius}.
The lower two panels show the $p$-values chosen by the binary search algorithm
for two selected values of $k$.

\section{Perturbative theory of fixed points at weak coupling} \label{app:perturb}

At small coupling $k$, a new set of fixed points emerge that we can account for
using a perturbative expansion of the averaged equations.

First, we motivate why we might expect more fixed points to exist. By applying
the averaging methods of the section stating the fixed point formula, it is
straightforward to show that \eqnref{eq:trimerfp} also applies to a single,
decoupled nonlinear Mathieu oscillator if we set $\omega = 1$.
\eqnref{eq:trimerfp} is also $\pi$-periodic in the phase $c$, meaning that
within the range of phases $[0,2\pi)$, there are in fact two physically
distinct phases, differing by $\pi$, that yield a solution. [While it may appear
that  \eqnref{eq:trimerfp} admits four fixed points in this range, two of
them are nonphysical as they lead to imaginary values of the amplitude.]  

When the trimer is completely decoupled, these considerations of an
independent Mathieu oscillator would indicate that there are eight
fixed points accessible to the system, one for each choice of the phases of the
three oscillators. Since the phase of the parametric drive determines the phases
of the fixed points, and the phase of the parametric drive differs by $2\pi/3$
between adjacent oscillators, the phase spaces of the oscillators are identical
except for a displacement of $2\pi/3$ in the phase direction. This means that
the fixed points can have phases that differ by $2\pi/3$ or by $\pi/3$ in the
limit of zero coupling, depending on the choice of each individual oscillator's
phase. 
The chiral steady state with phase offsets of $2\pi/3$ between adjacent oscillators accounts for two of these fixed points---identical up to an overall phase shift of $\pi$ due to the period-doubling of the modulation in the response.
The remain six fixed points involve phase spacing at or near $\pi/3$.
We conjecture  that if we introduce a very small coupling in the
trimer, these fixed points would persist as we increase the coupling
until a bifurcation occurred. Theory and numerical experiments, which we present
next, support the existence of these small-$k$, near-$\pi/3$-spaced fixed
points.

\begin{figure}
    \centering
    \includegraphics{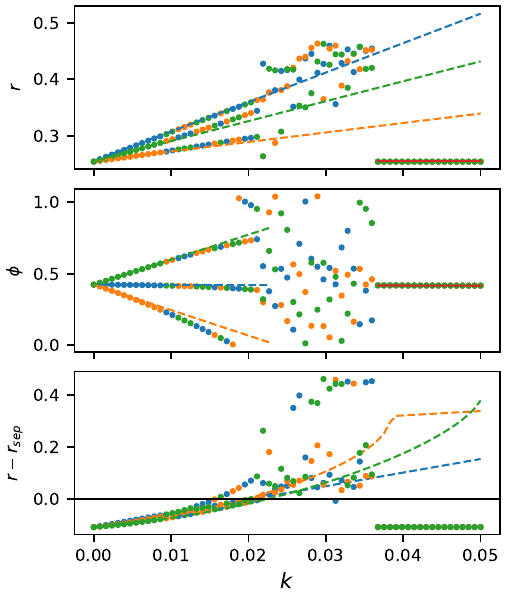}
    \caption{Perturbation theory of the small-$k$ fixed points. Top two panels: Steady-state amplitudes $r$ and phases $\phi$ of the trimer as
    functions of the coupling $k$ at small values of $k$. Each color indicates a
    different oscillator. Dots show numerical results; dashed curves show perturbation theory predictions [solutions to \eqnref{eq:csjdj}].
    Solid red line indicates the prediction for the $2\pi/3$-spaced fixed point (the chiral steady state). In the second plot, the
    phases are shown modulo $\pi/3$. Bottom panel:
    Difference between the numerical and perturbative amplitudes and the
    amplitude of the separating curve between open and closed undamped phase space curves at the corresponding phase value. 
    Parameters used are
    $\epsilon=0.01$ and $\delta = 0.1$; the modulation frequency $\gamma$ was
    varied with $k$ so that $\gamma = 2\sqrt{1+3k}$. The initial conditions used
    are $x_j(0) = 0.1\cos(\phi_j)$, $v_j(0) = -0.1\sin(\phi_j)$, where $\phi_1 =
    0$, $\phi_2 = \pi/3$, $\phi_3 = -\pi/3$; ie, the system was initialized with
    the oscillators separated in phase by $\pi/3$. The integration time interval
    was $[0,5000]$.}
    \label{fig:piover3}
\end{figure}

To derive the perturbative expressions for these fixed points, we insert an
ansatz with approximately $\pi/3$-spaced phases into equation
\eqref{eq:complexavgeoms} and expand to first-order in $k$. We start by
rearranging \eqnref{eq:complexavgeoms}. Letting $a = \delta/(4\omega)$,
$b=3/(8\omega)$, $g = k/(2\omega)$, and $U =
-\epsilon/2+i[k/\omega+(1-\omega^2)/(2\omega)]$, we have 
\begin{align}\label{eq:complexavgsmallk}
    (4b|A_j|^2-iU)A_j+a e^{i\beta_j}A_j^* - gi(A_{j-1}+A_{j+1})
\end{align}

Here, as before, $\beta_j = 2\pi(j-1)/3$ with $j \in \{1,2,3\}$ and index
arithmetic understood to be mod 3 (e.g., $j+3 = j$). Into this we substitute the
ansatz 
\begin{align}\label{eq:smallkansatz}
    A_1 &= \frac{r_1}{2}e^{ic_1} \\
    A_2 &= \frac{r_2}{2}e^{i(c_2+\pi/3)}\nonumber \\
    A_3 &= \frac{r_3}{2}e^{i(c_3-\pi/3)}\nonumber
\end{align}

In this ansatz, we take $r_j$ and $c_j$ to be perturbed by the coupling $k$
relative to their values when $k = 0$. In other words, we take 
\begin{align}\label{eq:ansatzrjcj}
    r_j &= r + k s_j \\
    c_j &= c + k d_j \nonumber 
\end{align}
where $r$ and $c$ are solutions to the averaged fixed-point equation for a
single, decoupled nonlinear Mathieu oscillator, which are the values given in
\eqnref{eq:trimerfp} with $\omega = 1$. Inserting \eqnref{eq:smallkansatz} with
\eqnref{eq:ansatzrjcj} into \eqnref{eq:complexavgsmallk}, setting the time
derivatives to zero, multiplying through by $\omega$, and expanding to first
order in $k$, we arrive at three equations. The terms in these equations that
are order 0 in $k$ cancel, since they reproduce the fixed point equation for an
isolated nonlinear Mathieu oscillator. 

Keeping terms to linear order in $k$, we obtain three decoupled
equations of the form 
\begin{align}\label{eq:csjdj}
    C(s_j,d_j) = D_j 
\end{align}
where 
\begin{align} 
    C(s,d) &= \frac{\delta}{4}e^{-2ic}(-ird+s) \nonumber \\
    &+ \frac{1}{8}\left[(-4r+6ir\epsilon) + (9r^2+4i\epsilon)s + (3ir^2-4r\epsilon)d\right] \nonumber
\end{align}
and the $D_j$ are
\begin{align}
    D_1 &= r\cos(\pi/3), \nonumber \\ 
    D_2 &= -ir\sin(\pi/3), \nonumber \\ 
    D_3 &= ir\sin(\pi/3). \nonumber 
\end{align}

From \eqnref{eq:csjdj}, we can obtain a linear equation in the two
perturbation variables $s_j$ and $d_j$ by equating the real and imaginary parts
on either side. Solving this equation for each value of $j$ and multiplying by
$k$ gives the perturbations to the amplitude and phase variables of the fixed points relative to their value in the decoupled system.
Our ansatz \eqnref{eq:smallkansatz} assigns zero relative phase to oscillator one; we obtain two more steady states with $\pi/3$ phase spacing by cyclically permuting the oscillator indices in the ansatz.
Additionally applying the overall $\pi$ phase shift allowed by the discrete
time-translation symmetry to each permutation generates the expected six fixed
points with $\pi/3$ phase differences.

In \figref{fig:piover3}, we compare the values of the predicted fixed points to
steady-state values obtained via numerical integration. For small values of $k$,
the amplitudes and phases agree with perturbation theory up to cyclic
permutations of the indices. Past a threshold $k$ value, the numerical features
diverge from the theory; this breakdown of perturbation theory corresponds
roughly to the value of $k$ at which the fixed points cross the heteroclinic
orbit between open and closed curves in the phase space projections for each
oscillator (\figref{fig:piover3}, bottom; see \appref{app:undamped} for details
of this heteroclinic orbit). Beyond a higher threshold coupling value, the
system converges to the $2\pi/3$-spaced fixed point---the chiral steady state
discussed in the main text.

At intermediate $k$ between the two threshold values, we observed quasi-steady states where the individual oscillators do not settle into a constant amplitude or phase, yet the trajectories exhibit some underlying regularity, reminiscent of beats arising from multiple harmonics.
These trajectories appear to be consistent with the self-oscillatory states described recently in Ref.~\onlinecite{Mahore2026a}, where a low-amplitude phase defect circulates persistently around the ring in a direction chosen by the modulation wave.
The states are expected when the coupling is so low that the modulation at twice the traveling-wave frequency $\omega=\sqrt{1+3k}$ also resonates with the fundamental mode at $\omega_0 = 1$. 
At high enough coupling, the parametric resonances no longer overlap, and the traveling-wave mode is uniquely excited by the modulation which causes the nonlinear trajectory to converge consistently to the $2\pi/3$-spaced chiral steady state as indicated by the collapse to the red line in \figref{fig:piover3}.
As we showed in the main text, the basin of attraction of the chiral steady state in the full phase space grows with coupling strength (\figref{fig:basin_radius}).

\end{document}